\documentclass[%
 reprint,
 superscriptaddress,
 amsmath,amssymb,
 aps,
 pra,
floatfix,
]{revtex4-2}

\usepackage[utf8]{inputenc}
\usepackage{amsmath, amssymb, mathtools, braket}
\usepackage{upgreek}
\usepackage{csquotes}
\usepackage{graphicx} 
\usepackage{dcolumn}% Align table columns on decimal point
\usepackage{bm}% bold math
\usepackage{xcolor}
\usepackage{booktabs} % \toprule / \midrule / \bottomrule
\usepackage{numprint} % to truncate + round digits in table
\usepackage{lmodern} % more font sizes
\usepackage{subfigure}

\usepackage{hyperref} % must be last
\usepackage[hyphenbreaks]{breakurl}
\hypersetup{
    pdftitle={Ab initio multichannel calculation of the Bethe surface and Compton defects for molecular hydrogen},
}

\newcommand{\mbf}[1]{\mathbf{#1}}
\newcommand{\mrm}[1]{\mathrm{#1}}
\newcommand{\op}[1]{\hat{\mathrm{#1}}}
\newcommand{\labeq}[1]{\label{eq:#1}}
\renewcommand{\refeq}[1]{Eq.\,(\ref{eq:#1})}
\newcommand{\labsec}[1]{\label{sec:#1}}
\newcommand{\refsec}[1]{Sec.\,\ref{sec:#1}}
\newcommand{\labfig}[1]{\label{fig:#1}}
\newcommand{\reffig}[1]{Fig.\,\ref{fig:#1}}
\newcommand{\labtbl}[1]{\label{tbl:#1}}
\newcommand{\reftbl}[1]{Table\,\ref{tbl:#1}}
\newcommand{\rme}{\mathrm{e}}
\newcommand{\rmd}{\mathrm{d}}
\newcommand{\im}{\mathrm{i}}

\DeclarePairedDelimiter\abs{\lvert}{\rvert}
\newcommand{\psich}{\psi^{(-)}_{a\, \mbf p_a'}}
\newcommand{\psiche}[1][E]{\psi^{(\Gamma,-)}_{#1',\, a\lambda}}

\graphicspath{{./scripts}}

\begin{document}

%\preprint{APS/123-QED}

\title{\texorpdfstring{\emph{Ab initio}}{Ab initio} multichannel calculation of the Bethe surface and \texorpdfstring{\\}{} Compton defects for molecular hydrogen}
\author{Hakon Volkmann}
\email{hakon.volkmann@physik.hu-berlin.de}
\affiliation{AG Moderne Optik, Institut f\"ur Physik, Humboldt-Universit\"at zu Berlin, Newtonstr. 15, 12489 Berlin, Germany}
\author{Alejandro Saenz}
\email{alejandro.saenz@physik.hu-berlin.de}
\affiliation{AG Moderne Optik, Institut f\"ur Physik, Humboldt-Universit\"at zu Berlin, Newtonstr. 15, 12489 Berlin, Germany}

\date{\today}

\begin{abstract}
  A correlated multichannel computation of the generalized oscillator-strength density for electrons scattering off randomly oriented molecular hydrogen at the internuclear separation of $1.4$ $a_0$ is being presented.
  The numerical procedure features a novel free-boundary method based on configuration interaction of ionic orbitals expanded in B splines using prolate-spheroidal coordinates.
  Satisfactory agreement between present and experimental outcomes has been found, highlighting the importance of final-state correlation in the description of the single-ionization continuum.
  The systematic and robust convergence behavior of the method further allowed for assessing the validity of the binary-encounter approximation, providing a confirmation of experimentally determined Compton defects for H$_2$ using a single consistent theoretical description.
  Overall, the present findings are expected to be of relevance for, \emph{e.g.}, time-resolved inelastic photon scattering in ultrafast molecular-structure imaging or for calbibrating electron-impact energy-loss measurements on molecular tritium, as required for the KATRIN neutrino mass experiment.
\end{abstract}

\maketitle

\section{Introduction}

Single ionization of H$_2$ molecules by fast electron impact is fundamental process, which has been studied both theoretically and experimentally beginning in the 1960s~\cite{gosd:boer61, gosd:geig64} and remained subject of extensive research efforts throughout the 1970s~\cite{gosd:ulsh74, gosd:lee77, gosd:bonh73, gosd:liu73} and 1980s~\cite{gosd:leun83, gosd:bonh86, gosd:zura88, gosd:cher89}.
The relative but yet non-trivial simplicity of H$_2$ renders it an ideal benchmark system for studying the interplay between various approximations and molecular properties, including electron correlation, nuclear motion, or nonspherical multicenter effects.
Concomitantly, being the most-abundant molecule in stellar and gaseous planetary ionosphere environments, quantitative understanding of electron-impact ionization processes also has practical applications in, \emph{e.\,g.}, plasma~\cite{gosd:yoon10} or astrophysics~\cite{gosd:chad16}.

If the impacting electrons are sufficiently fast, the theoretical treatment of the scattering process can be entirely pursued within the non-relativistic first Born approximation (FBA), as effects due to exchange or relativistic kinematics are, to a large extent, negligible.
Furthermore, if the kinetic energy of the ejected target electron (and thus the transferred momentum) is significantly larger than the binding energy, its interaction with the residual ion may be neglected upon collision, giving rise to the binary-encounter approximation (BEA)~\cite{gosd:inok71, gosd:bonh73, gosd:kim94}.
As is well-known, this reasoning also applies to Compton scattering~\cite{gosd:prat10,gosd:liu14,gosd:zhao15}, where 
the connection to the impulse approximation (IA)~\cite{gosd:eise70} is established by the use of plane waves for the modelling of the ejected electron.
The driving reason for measuring Compton profiles in molecules and lattices up until recently~\cite{gosd:wang20,gosd:onit22,gosd:wan23} is its connection to the initial-state momentum distribution of the target, providing a mean for orbital imaging~\cite{gosd:star01, gosd:carr19, gosd:roma26} or an alternative tool to experimentally assess the quality of theoretically determined wavefunctions beyond their effect on the initial-state energy~\cite{gosd:eise70}.
Alternatively, $(e, 2e)$ coincidence measurements on ejected and scattered electrons yield information on the initial-state momentum distribution of the target in a similar manner, as is done in electron momentum spectroscopy (EMS) experiments~\cite{gosd:mcca91, gosd:neud99, gosd:taka09, gosd:zhan14}.

In order to assert the validity of the binary-encounter and impulse approximations~\cite{gosd:kapl03}, recent high-precision EMS experiments for Ne atoms~\cite{gosd:naka25} and measurements of Compton profiles~\cite{gosd:barl78, gosd:ruec78} for He and H$_2$ have been undertaken.
Here, the latter results will serve as an ideal benchmark for assessing the accuracy of the present theoretical data for molecular hydrogen.
Furthermore, limited computational capabilities only allowed for matching individual facets of the experimental data to theory if different descriptions are being adopted simultaneously.
These, however, occasionally even show a rather strong mutual disagreement~\cite{gosd:zura88} beyond their \emph{a priori} unknown domains of applicability.
In order to alleviate these discrepancies, the quality of numerical treatments needs to be enhanced beyond what was possible in the past~\cite{gosd:smit77, gosd:smit80, gosd:zura88}, ideally giving rise to a single unified description without being forced to rely on BEA, Hartree-Fock, Coulomb, or distorted-wave approximations within the FBA picture.

Recently, the demand for accurate electron-impact ionization cross sections of H$_2$ isotopologues reemerged from an entirely different direction within the context of the KATRIN neutrino-mass experiment~\cite{gosd:aker25}.
Its primary goal is to determine the absolute mass of electron-antineutrino by fitting the measured emission rate of $\upbeta$ electrons to the high-energy limit of the mass-parametrized theory~\cite{gosd:klee19}.
The $\upbeta$ electrons originate (along with the antineutrinos) from the decay of tritium T$_2$ molecules within the tritium source, which, due to the high sensitivity required in order to accurately perform the fit, needs to contain gaseous tritium at a sufficiently high density.
However, due to this abundance of target molecules, the $\upbeta$ electrons may scatter once or multiple times with other T$_2$ molecules within the source before reaching the detector.
As such, the energy loss suffered by the $\upbeta$ electrons distorts the measured energy spectrum, hence adding a systematic source of error.
In order to remedy this effect, a sufficiently accurate energy-loss spectrum is required, which can be produced experimentally by employing an electron gun that is calibrated using a theoretical model~\cite{gosd:aker21}.
Surprisingly, although the process itself has been the subject of research that dates back more than six decades ago~\cite{gosd:boer61}, data specifically for ionization processes spanning a sufficiently large range of energy and momentum transfers that would allow for a sufficiently accurate computation of the energy-loss spectrum does seemingly not exist.
Instead, the focus of theoretical research shifted towards the description of triply-differential cross sections and second Born effects~\cite{gosd:zura88, gosd:lahm88, gosd:weck01, gosd:houa03, gosd:stia03, gosd:sero05, gosd:capp06, gosd:sero09, gosd:yuro14, gosd:dhan20} and beyond that became increasingly available from intricate coincidence measurements in EMS and other $(e, 2e)$ experiments~\cite{gosd:leun83, gosd:cher89, gosd:li18}.
Whereas extensive electron-impact excitation data is readily available for H$_2$~\cite{gosd:arri80, gosd:kolo82, gosd:liu93, gosd:borg99, gosd:xu18}, HD, and D$_2$ isotopologues~\cite{gosd:kolo83,gosd:liu23}, properly describing the infinitely degenerate multichannel ionization-continuum poses a more involved challenge for both, theoretical and computational models.
As such, theoretical work~\cite{gosd:zura88} on impact ionization beyond binary-encounter or semi-empirical binary-encounter dipole models~\cite{gosd:rudd91} has remained scarce or restricted to total ionization cross sections~\cite{gosd:liu87}.
In the following years, more elaborate computational $L^2$ methods capable of computing such scattering wavefunctions have been developed~\cite{gosd:mart91, gosd:bros92, gosd:bros92a, gosd:sanc97, gosd:mart99}, but were seemingly applied predominantly for ionization processes induced by (few-)photon absorption~\cite{gosd:toff16, gosd:mara17, gosd:toff24} rather than by electron impact.

A rather elegant workaround for directly determining a numerical continuum representation for the ejected electrons involves the combination of the Kramers-Kronig (or dispersion) relation on the resolvent in connection with the complex-scaling method~\cite{gosd:agui71, gosd:bals71, gosd:simo72, gosd:simo73}.
Bypassing the tedious labour of explicitly analyzing the multichannel structure of the target~\cite{gosd:froe85}, it essentially allows for obtaining the doubly-differential scattering cross-section (DDCS) in an alternative way.
While this approach appears to be promising, preliminary results involving electron scattering so far only have been computed for helium atoms~\cite{gosd:saen93,gosd:saen96} and could not be converged sufficiently far into the regions of larger scattering angles required for the KATRIN experiment.
In order to fill this gap, this work approaches the scattering problem head-on and presents an \emph{ab initio} approach based on a configuration-interaction (CI) method using ionic H$_2{}^+$ orbitals constructed with B splines~\cite{gosd:bach01} in prolate spheroidal coordinates.
Within the Born-Bethe theory, the generalized oscillator-strength density (GOSD) is computed by combining a ground-state description, featuring a large number of configurations with an accurate, multichannel model of the scattering continuum which combines CI with a recently developed free-boundary (FB) method.

The contents of this work may be outlined as follows.
In \refsec{theory}, all observables and the mathematical framework required for the description of bound and scattering states is summarized.
Starting with scattering cross-sections in \refsec{cs}, the discussion is followed by the partial-wave analysis in \refsec{pw} and the introduction of the CI basis in \refsec{ci}.
Afterwards, in \refsec{comp}, computational details on the bound-state and free-boundary computations are given in Secs. \ref{sec:bound} and \ref{sec:fbm}, respectively, followed by a discussion of the selection rules in \refsec{form}.
The results which feature the Bethe surface as well as the Compton profile are presented in \refsec{results} and successively discussed in \refsec{disc}.
Finally, a summary can be found in \refsec{summary}.

\section{Theory}\labsec{theory}

In this section, if not stated otherwise, the atomic unit system with $\hbar=m_e=4\pi\epsilon_0=1$ is being adopted.
Throughout this work, the target's four-body center of mass (COM) is assumed to approximately coincide with the two-body \emph{nuclear} COM, which is justifiable by the large mass asymmetry between the nuclei and electrons.
As such, the discussion is entirely performed within the total COM system of projectile and target molecule.

\subsection{Cross sections}\labsec{cs}

The GOSD, denoted by $\rmd f/\rmd E$, is related to the orientation-averaged, single-ionization DDCS by~\cite{gosd:bonh86}
\begin{equation}
 \labeq{egosd}
  \frac{\rmd f(K,E)}{\rmd E} \;=\; \frac{E\, K^2}{2}\, \frac{p_i}{p}\, \left\langle\frac{\rmd^2\sigma}{\rmd \Omega\, \rmd E}\right\rangle_{\hat{\mbf R}}\,,
\end{equation}
with $E$ denoting the energy transfer (or loss) of the projectile to the target, $p_i$ the modulus of the incident and $p$ that of the scattered electron's momentum relative to the COM as well as $\Omega$ the scattering angle.
The definition given in \refeq{egosd} further neglects all corrections due to spin exchange and relativistic effects, the former being justified if the incident momentum is sufficiently large for the projectile to be effectively distinguishable.
However, note that the two target electrons are still being treated as indistinguishable for both initial and final states.
The modulus of momentum transfer $\mbf K = \mbf p_i - \mbf p$ is specified by the polar scattering angle $\vartheta_p$ via
\begin{equation}\labeq{angle}
  \rmd\vartheta_p \;=\; \frac{K}{p_i\,p}\,\rmd K
\end{equation}
with $\rmd\Omega=\sin(\vartheta_p)\, \rmd\vartheta_p\,\rmd\varphi_p$.
Introducing the notation $\{\mbf P\} = \left(\mbf p,\mbf p_a',\mbf p_i\right)$ with ejected-electron momentum vector $\mbf p_a'$,
the orientation-averaged DDCS for scattering an electron incident on the molecule in channel $i$ into channel $a$ is given by
\begin{equation}
 \labeq{orient}
  \left\langle\frac{\rmd^2\sigma_{i \,\to\, a}}{\rmd\Omega\, \rmd E}\right\rangle_{\hat{\mbf R}} =\; \frac{1}{4\pi}\,\frac{p \, p_a'}{p_i} \iint \abs{F_{i\,\to\,a}(\{\mbf P\};\mbf R)}^2\,\rmd \hat{\mbf R}\,\rmd\Omega'\,,
\end{equation}
where the adiabatic scattering amplitude $F_{i\,\to\,a}(\{\mbf P\};\mbf R)$ is integrated over the direction $\Omega'=(\vartheta_{p_a'},\varphi_{p_a'})$ of the ejected electron momentum as well as the orientation of the molecular axis $\mbf R=R\, \hat{\mbf R}$ with internuclear distance $R$.
Summing over all (unresolved) \emph{open} final channels $a$ for a given energy loss $E$ results in the total DDCS used in \refeq{egosd},
\begin{equation}\labeq{sum_chan}
  \left\langle\frac{\rmd^2\sigma}{\rmd\Omega\, \rmd E}\right\rangle_{\hat{\mbf R}} \;=\;\;\; \mathclap{\,\displaystyle\int\limits_{I_{a} < E}}\mathclap{\displaystyle\sum}\;\;\;\left\langle\frac{\rmd^2\sigma_{i \,\to\, a}}{\rmd\Omega\, \rmd E}\right\rangle_{\hat{\mbf R}}\,,
\end{equation}
where the initial channel label $i$ is implied on the l.\ h.\ s.
Adopting the Born-Oppenheimer (BO) approximation, the label $i$ ($a$) shall identify a BO state of the (ionized) H$_2$ (H$_2{}^+$) target of energy $\epsilon_{n_i\nu_i}$ ($\epsilon_{n_a\nu_a}$) with $n_i$ ($n_a$) indicating the electronic potential and $\nu_i$ ($\nu_a$) the vibronic excitation within that potential, resulting in a threshold energy $I_a=\epsilon_{n_a\nu_a}-\epsilon_{n_i\nu_i}$.
All the target states are ordered by ascending values of $I_a$, whereas the integral symbol in \refeq{sum_chan} is given to emphasize the continuum of vibronic dissociation channels being included as well.
Note, for a given value of $E$, conservation of energy requires for the modulus of $\mbf p_a'$ to fulfill
\begin{equation}\labeq{eloss}
  p_a' \;=\; \sqrt{2(E-I_a)}
\end{equation}
with
\begin{equation}
  E \;=\; \frac{p_i^2 - p^2}{2}\,.
\end{equation}

If an ensemble of molecules is assumed to be cooled down to sufficiently low temperatures such that there remains no appreciable population of higher electronic or vibronic states, the initial channel $i$ can be taken as the lowest vibronic state $\nu_{0}(R)$ of the electronic ground state $\psi^{\left(^1\Sigma_\mrm{g}^+\right)}_0(\mbf r_1,\mbf r_2;R)$ (with electronic coordinates $\mbf r_j,\ j=1,2$), if the BO approximation is applied.
Adopting the latter is well justifiable by the high incident energy and the correspondingly short collision times being significantly shorter than the molecular vibration or rotation periods, hence providing a stroboscopic view on the target molecular motion upon scattering.
The omission of rotational states from both the initial and final channels is a consequence of the orientation averaging, which is well known to be equivalent~\cite{gosd:iiji63, gosd:lane80, gosd:kolo82, gosd:read63, gosd:cart67, gosd:arri90, gosd:weck00} to taking a (thermal) ensemble average over initial states and summing over all final rotational channels, if the (small) rotational energies are ignored and scattering does not excite high-lying rotational states~\cite{gosd:cart67}.

In the first Born approximation, which is assumed to be valid due to the large magnitude of incident momentum $p_i$, the incident and scattered electron's wavefunction is approximated by plane waves, the length-form scattering amplitude thus being given by
\begin{equation}
 \labeq{sc_amp}
 F_{i\,\to\, a}(\{\mbf P\};\mbf R) \,=\, -\frac{2}{K^2}\; \sum\limits_{j=1}^2\Braket{\nu_a\,\psich | \rme^{\im\mbf K\cdot\mbf r_j} | \nu_i\,\psi_{i}}\,,
\end{equation}
which is the commonly known result of Born-Bethe theory~\cite{gosd:beth30, gosd:inok71}.
Here, $\psich(\mbf r_1,\mbf r_2;R)$ denotes the \enquote{half-scattering} state with incoming boundary conditions that describes the ejected electron, departing with momentum $\mbf p_a'$ from the molecular ion that is left behind in electronic state $\phi_a$ and (potentially dissociative) vibronic state $\nu_a$.

Finally, once the GOSD in \refeq{egosd} is known, it can be further utilized to compute an effective Compton profile according to
\begin{equation}
  \labeq{compprof}
  J(K,q) \;=\; \frac{K^3}{2\,E}\, \frac{\rmd f(E,K)}{\rmd E}\,,
\end{equation}
with $q = E/K \,-\, K/2$.
In the limit of $K$ being large relative to the average momentum of the bound electrons, both the BEA and the IA yield the Compton profile in terms of the orientation-averaged initial-state momentum density $\rho(p)$~\cite{gosd:vrie68, gosd:eise70, gosd:zura88, gosd:zhao15}, 
\begin{equation}
  \labeq{beacomp}
  J^\mrm{(BEA)}(q) \;=\; 2\pi\int_{\abs q}^\infty p\,\rho(p)\,\rmd p\,,
\end{equation}
with the notable independence of $K$, compared to \refeq{compprof}, rendering it a truly projectile-independent property of the target system.
For transferred momenta where the BEA is not sufficiently accurate, both the height of the Compton profile peak and its position's deviation from $q=0$, commonly known as the Compton defect~\cite{gosd:barl78,gosd:ruec78,gosd:froe85,gosd:bell86}, are particularly sensitive to the initial and final state description and thus will serve as a benchmark for the quality of the present results.

\subsection{Electronic partial waves}\labsec{pw}

In order to compute the scattering amplitude \refeq{sc_amp} and, eventually, the full GOSD \refeq{egosd}, the channel-resolved scattering states $\psich$ have to be determined.
For this purpose, it is convenient to work within a partial-wave framework, as it allows for obtaining a simple closed-form expression for the orientation integral \refeq{orient} that saves one from the efforts of computing scattering wave functions for all but one molecular orientation.

The original code~\cite{gosd:vann04} (which ultimately dictates the here chosen \emph{ansatz} for the actual computations) has been written in terms of (two-center) prolate spheroidal coordinates $(\xi,\eta,\varphi)$, which are related to the Cartesian coordinate system with common origin by
\begin{align}
  \begin{split}
   \labeq{psc}
    x \;&=\; \frac{R}{2}\cos(\varphi)\, \sqrt{(\xi^2-1)\, (1-\eta^2)} \\
    y \;&=\; \frac{R}{2}\sin(\varphi)\, \sqrt{(\xi^2-1)\, (1-\eta^2)} \\
    z \;&=\; \frac{R}{2}\,\xi\eta\,.
  \end{split}
\end{align}
Here, $\varphi$ is the same azimuth angle as is used in spherical coordinates.
As such, the partial-wave description has to take place in terms of the (orthonormal) \emph{spheroidal} harmonics
\begin{equation}
  \labeq{sph_harm}
  \Upsilon^m_\lambda(\eta,\varphi\,;\, C_a') \;=\; \frac{\mathcal N^{|m|}_{\lambda}}{\sqrt{2\pi}}\, S^{|m|}_{\lambda}(\eta\,;\, C_a')\,\rme^{\im m\varphi}\,,
\end{equation}
with $C_a'=p_a'R/2$, normalization $\mathcal N^m_\lambda$ (following the convention of Chu and Stratton~\cite{gosd:flam57}), and angular spheroidal wavefunction $S^{|m|}_\lambda$~\cite{gosd:flam57}.
The latter shares, though not all, but many of the numerous useful properties of the more common spherical harmonics $Y^m_\ell(\vartheta,\varphi)$ given in (single-center) spherical coordinates.
In particular, owing to their orthonormality property, it is still possible to carry out the orientation-averaging in an efficient closed-form manner.

With above considerations in mind, the scattering state $\psich$ is expanded as~\cite{gosd:sero02, gosd:miya12}
\begin{align}
 \begin{split}
  \labeq{psich}
  \psich(\mbf r_1,&\mbf r_2;R) \,=\, \sqrt{\frac{2}{\pi\,p_a'}}\, \sum\limits_{m=-\infty}^\infty \sum_{\lambda=|m|}^\infty \im^{\lambda}\rme^{-\im\sigma^{|m|}_{\lambda}}\,\times\\
  \times\, \Upsilon^{m^*}_{\lambda}&\left(\cos\vartheta_{p'_a},\varphi_{p'_a}\,;\, C_a'\right)\, \psi^{(\Gamma_{a \lambda m},-)}_{E',\, a\lambda}(\mbf r_1,\mbf r_2; R)\,,
 \end{split}
\end{align}
introducing the partial-wave component $\psi^{(\Gamma_{a\lambda m},-)}_{E',\,a \lambda}(\mbf r_1,\mbf r_2; R)$ of molecular symmetry $\Gamma_{a\lambda m}$, which is unambiguously determined by the single-particle symmetry $\gamma_a$ of the ionic channel labelled by $a$ and the summation indices $m$ and $\lambda$, \emph{cf.}\ \refeq{sym} in appendix \ref{sec:sym}.
Furthermore, the two-center Coulomb scattering phase shift is denoted as $\sigma^{|m|}_{\lambda}$~\cite{gosd:sing23, gosd:volk26} and
\begin{equation}\labeq{eprime}
 E'(R)=p_a'{}^2/2+\epsilon^{\gamma_a}_{n_a}(R) 
\end{equation}
is the ionic channel's potential curve, shifted up by the kinetic energy of the ejected electron.
In fact, the states $\psiche[E]$ are improper eigenstates (see  \refeq{tese} below) of the electronic target Hamiltonian at fixed internuclear separation $R$ for the given symmetry $\Gamma$ and energy $E'$, whose numerical computation details are subject of \refsec{fbm}, but have to be properly characterized first.
In particular, they are required to fulfill the adiabatic asymptotic boundary conditions~\cite{gosd:star23}
\begin{align}
\begin{split}
  \labeq{mc_asymp}
  \psi&^{(\Gamma,-)}_{E',\,a\lambda}(\mbf r_1,\mbf r_2;R) \underset{\xi_1\to\infty}{\sim} \op{\mathcal A}^{(S)} \sum\limits_{b=1}^{n_c} \sum_{\lambda'=|\bar m_b|}^\infty \frac{-\im}{R\, \xi_1 \sqrt{p'_b}}\, \times\\
  &\times\, \phi^{\gamma_b}_{n_b}(\mbf r_2;R)\,\Upsilon^{\bar m_b}_{\lambda'}(\eta_1,\varphi_1\,;\, C_b')\,\times\\
  &\times\, \bigg[\delta_{ab}\delta_{\lambda\lambda'}\,\exp\left(\im\theta^{|\bar m_b|}_{b\lambda'}\right) \,- \,S^{(\Gamma)\dagger}_{a\lambda,b\lambda'}\,\exp\left(-\im\theta^{|\bar m_b|}_{b\lambda'}\right)\bigg]\,,
\end{split}
\end{align}
where the first sum runs over all $n_c$ \emph{adiabatically} open ionic channels with label $b$ (to be explained below) and the components of the $S$ matrix being denoted as $S^{(\Gamma)}_{a\lambda,b\lambda'}$.
The second sum runs over all degenerate pseudo-angular channels labelled by $\lambda'$, which are coupled with one another just as the ionic channels are.
Furthermore, in \refeq{mc_asymp}, the phase
\begin{equation}
  \labeq{sph_phase}
  \theta^{|\bar m_b|}_{b\lambda'} \,=\, C'_b\,\xi_1 - \frac{\lambda'\pi}{2} + \frac{1}{p'_b}\ln(2\,C'_b\,\xi_1) + \sigma^{|\bar m_b|}_{\lambda'}
\end{equation}
is given in terms of the channel momentum
\begin{equation}\labeq{pbp}
    p'_{b}(R) \,=\, \sqrt{2\, E'(R)-2\, \epsilon^{\gamma_b}_{n_b}(R)}
\end{equation}
for the ionic molecular-state potential curve $\epsilon^\gamma_{n_b}(R)$.
The precise meaning of what is being called here an \enquote{adiabatically open} ionic channel becomes transparent by \refeq{pbp}.
For a given value of $R$, there is a (generally $R$-dependent) number of ionic channels $n_c(R)$ for which $E'(R)>\epsilon^{\gamma_b}_{n_b}(R)$.
At first glance, it might appear surprising that this number $n_c$ generally is quite different from the \emph{actual} number of open channels for a given energy loss of the projectile as being given in \refeq{sum_chan}.
This, however, should be interpreted as a mere artifact of the adiabatic description.
The final BO state after collision, as required from \refeq{sc_amp}, is $\nu_a(R)\psich(\mbf r_1,\mbf r_2\,;R)$, whose  asymptotic interpretation is that of an ejected, hence \emph{outgoing} electron (say, at $\mbf r_1$) with momentum $\mbf p'_a$, leaving behind a residual ion in electronic state $\phi^{\gamma_a}_{n_a}(\mbf r_2\,;R)$ and vibronic state $\nu_a(R)$.
This state carries the energy (\emph{cf.}\ \refeq{eloss})
\begin{equation}
  \epsilon_{n_a\nu_a}+p_a'{}^2/2=\epsilon_{n_i\nu_i}+E\,,
\end{equation}
\emph{i.e.}, that of the initial target state energy plus the energy $E$ transferred from the projectile.
In the adiabatic picture, the scattering takes place when the molecule is stretched to a certain instantaneous value of $R$.
The adiabatic energy $E'(R)$, \emph{cf.}\ \refeq{eprime}, for the scattering state $\psiche[E]$ is required to represent an outgoing electron of definite momentum in channel $a$ \emph{at this} value of $R$ and, therefore, may take any value necessary.
Depending on $n_c(R)$, different numbers of ionic channels $b$ are adiabatically open at this energy and, thus, coupled.
Ultimately, presence (or absence) of these adiabatically open channels, however, has no direct influence whatsoever on the scattering amplitude \refeq{sc_amp} itself, as they are asymptotically only contributing to the \emph{incoming} waves (\emph{cf.}\ \refeq{mc_asymp}) with channel momentum $p'_b(R)$, \refeq{pbp}, whereas the \emph{only} outgoing contribution of $\psiche[E]$ lies in the $a$ channel, which, by definition of $E'(R)$ \refeq{eprime}, is always adiabatically open.

The magnetic quantum number $\bar m_b$ in \refeq{mc_asymp} is uniquely determined by the total magnetic quantum number $\Lambda$ of the symmetry $\Gamma$ and the single-particle symmetry $\gamma_b$ of the channel, such that $\Lambda=|m_b+\bar m_b|$.
Likewise, for total parity $\mathcal P$ and ionic parity $\wp_b$, the admissible values for $\lambda'$ in the second summation of \refeq{mc_asymp} are even for both $\mathcal P$ and $\wp_b$ being \emph{gerade} or \emph{ungerade} and odd otherwise.
Finally, the $\op{\mathcal A}^{(S)}$ operator denotes the (anti-)symmetrizer, depending on the total electronic spin multiplicity given by the two-particle symmetry $\Gamma$ and the single-particle symmetry $\gamma$ (\emph{cf.}\ appendix \ref{sec:sym}).
In this work, only singlet states and, for $\Lambda=0$, states of $\Sigma^+$ reflection symmetry are considered.

Similar to the well-known plane-wave expansion in terms of spherical Bessel functions, the exponential expression in the form factor of \refeq{sc_amp} can be expanded as~\cite{gosd:flam57, gosd:liu93, gosd:sero02, gosd:sero05, gosd:sero09, gosd:tapl18} 
\begin{align}
\begin{split}
  \labeq{prosphexp}
  \rme^{\im\mbf K\cdot\mbf r_i} \;=\; \sum\limits_{L=0}^\infty\sum\limits_{M=-L}^L \Upsilon^{M*}_{L}(\cos\vartheta_K,\varphi_K\,;\,C)\,\Theta^M_{L}(\mbf r_i \,;\, C)\,,
\end{split}
\end{align}
with $C=KR/2$ and
\begin{equation}\labeq{expterm}
  \Theta^M_{L}(\mbf r_i\, ;\, C) \;=\; 4\pi\,\im^L\, \Upsilon^{M}_{L}(\eta_i, \varphi_i\,;\, C)\, R^{|M|}_{L}(\xi_i\,;\, C)\,,
\end{equation}
where $R^{|M|}_L$ denotes the radial prolate spheroidal wavefunction~\cite{gosd:flam57}.
Inserting Eqs. (\ref{eq:prosphexp}) and (\ref{eq:psich}) into \refeq{sc_amp} and subsequently into \refeq{orient}, one can perform both the orientation averaging and the $\Omega'$ integrals by the virtue of the orthonormality property of the spheroidal harmonics, yielding
\begin{align}
 \begin{split}
  \labeq{avg_cross_sec}
   &\left\langle\frac{\rmd\sigma_{i\,\to\,a}}{\rmd\Omega\,\rmd E}\right\rangle_{\hat{\mbf R}} \;=\; \frac{2}{\pi^2\,K^4}\,\frac{p}{p_i} \sum\limits_{\substack{m=-\infty\\\lambda=|m|}}^\infty \sum\limits_{\substack{M=-\infty\\L=|M|}}^\infty \abs*{A^{M(-)}_{L,\,a\lambda m}}^2\,,
 \end{split}
\end{align}
with the amplitudes being given as
\begin{equation}
 \labeq{ampl}
  A^{M(-)}_{L,a\lambda m} \;=\; \sum_{i=1}^2\Braket{\nu_a\,\psi^{(\Gamma_{a\lambda m},-)}_{E',\,a\lambda} | \Theta^M_{L}(\mbf r_i\,;\, C) | \nu_i\,\psi_i}\,.
\end{equation}
This can be seen to be formally identical to the result of the single-center expansion found in Ref.~\cite{gosd:zamm17}.
As is the case for the channel summation over $\lambda'$ in \refeq{mc_asymp}, the summation w.\,r.\,t.\ $L$ in \refeq{avg_cross_sec} can be restricted to even or odd values, depending on the total parity determined by the channel and $\lambda$.
Furthermore, by the selection rule \refeq{mselect} further laid out in \refsec{form}, the terms \refeq{ampl} for $M$ and $\lambda$ are only contributing for $M=\lambda$, and, for $|M|>0$, give an identical contribution for $\pm M$, hence enabling to restrict the summation accordingly.

\subsection{CI basis}\labsec{ci}

If a CI basis expansion is adopted for $\psiche[E]\,$, the integrals in \refeq{avg_cross_sec} can be computed as the sum of products of separate one-dimensional integrals for each particle and coordinate.
Thus, as before, let $\phi^{\gamma}_{n}(\mbf r;R)$ denote the ionic wavefunction that solves the single-electron problem
\begin{equation}
 \labeq{oese}
  \op{H}^\mrm{\left(H_2{}^+\right)}(R)\,\phi^{\gamma}_n(\mbf r;R) \;=\; \epsilon^\gamma_n(R)\,\phi^{\gamma}_n(\mbf r;R)\,,
\end{equation}
where $\epsilon^\gamma_n(R)$ denotes the electronic potential curve of the molecular ion.
The Hamiltonian in \refeq{oese} is
\begin{equation}
  \labeq{sham}
  \op{H}^\mrm{\left(H_2{}^+\right)}(R) \;=\; -\frac{1}{2}\, \Delta \,+\, \frac{1}{\abs*{\hat{\mbf r}+\mbf R/2}} \,+\, \frac{1}{\abs*{\hat{\mbf r}-\mbf R/2}}\,,
\end{equation}
which renders \refeq{oese} separable if expressed in prolate spheroidal coordinates, \emph{cf.}\ \refeq{psc}.
Consequently, the solutions can be written down as the Lam{\'e} product
\begin{equation}
  \labeq{orb}
  \phi^\gamma_n(\mbf r;R) \;=\; \frac{1}{\sqrt{2\pi}}\, X^{\abs{\gamma}}_n(\xi;R)\, Y^{\abs{\gamma}}_n(\eta;R)\,\rme^{\im m\varphi}\,,
\end{equation}
expressed in the body frame, which aligns $\hat{\mbf R}$ along the $\hat{\mbf z}$ axis and $\abs{\gamma}=(\abs{m},\wp)$.

Once obtained, two single-particle solutions $\phi^\gamma_n$ and $ \phi^{\bar\gamma}_{\bar n}$ are combined into a properly symmetry-adapted two-electron configuration using the molecular symmetry adaptor $\op{\mathcal S}^{(\Gamma)}$,
\begin{equation}
 \labeq{adapter}
  \Phi^{(\Gamma)}_{n\bar n}(\mbf r_1,\mbf r_2) \;=\; \op{\mathcal S}^{(\Gamma)} \left[\phi^\gamma_n(\mbf r_1)  \phi^{\bar\gamma}_{\bar n}(\mbf r_2)\right]
\end{equation}
(with the parametric $R$ dependence understood), which are given by
\begin{subequations}
 \begin{equation}
 \labeq{conf1}
 \Phi^{(\Gamma\notin\Sigma)}_{n\bar n}(\mbf r_1,\mbf r_2) \;=\; \frac{\phi_n^\gamma(\mbf r_1)\phi_{\bar n}^{\bar\gamma}(\mbf r_2) + (-1)^S\phi_{\bar n}^{\bar\gamma}(\mbf r_1)\phi_n^{\gamma}(\mbf r_2)}{\sqrt{2(1+\delta_{n\bar n}\delta_{\gamma\bar\gamma})}}
\end{equation}
for non-$\Sigma$ states and
\begin{align}
 \begin{split}
 \labeq{conf2}
 \Phi^{(\Gamma\in\Sigma^\pm)}_{n\bar n}(\mbf r_1,&\mbf r_2) \;=\; \frac{1}{\sqrt{4(1+\delta_{n\bar n}\delta_{\abs{\gamma}\abs{\bar\gamma}})(1+\delta_{0m}\delta_{0\bar m})}}\,\times\\
 &\times\,\Big[\phi_n^\gamma(\mbf r_1)\phi_{\bar n}^{\bar\gamma}(\mbf r_2) + (-1)^S\phi_{\bar n}^{\bar\gamma}(\mbf r_1)\phi_n^{\gamma}(\mbf r_2)\,+\\
 &\pm\, \left\{\phi_n^\gamma(\mbf r_1)\phi_{\bar n}^{\bar\gamma}(\mbf r_2) + (-1)^S\phi_{\bar n}^{\bar\gamma}(\mbf r_1)\phi_n^{\gamma}(\mbf r_2)\right\}^*\Big]
 \end{split}
\end{align}
\end{subequations}
for $\Sigma^\pm$ states, respectively.
Both, the initial ($\psi_i$) and partial-wave scattering states ($\psiche[E]$, \emph{cf.}\ \refeq{psich}) are then approximated by a finite linear combination of configurations Eqs.~(\ref{eq:conf1}, \ref{eq:conf2}),
\begin{equation}
 \labeq{ci}
  \psi^{(\Gamma)}(\mbf r_1,\mbf r_2;R) \;=\; \sum_{n\bar n} C^{(\Gamma)}_{n\bar n}\,\Phi^{(\Gamma)}_{n\bar n}(\mbf r_1,\mbf r_2;R)\,,
\end{equation}
with coefficients $C^{(\Gamma)}_{n\bar n}$ to be determined numerically by applying the Galerkin method. For further details on the overall approach, the reader is referred to Ref.~\cite{gosd:vann04}.

\section{Computational details}\labsec{comp}
\subsection{Bound states}\labsec{bound}

Both separable solutions $X^{\abs\gamma}_n(\xi)$ and $Y^{\abs\gamma}_n(\eta)$  (\emph{cf.}\ \refeq{orb}) are expanded in terms of respective B-spline bases,
\begin{subequations}
\begin{equation}
 \labeq{xdef}
  X^{\abs\gamma}_n(\xi) \;=\; (\xi^2-1)^{\abs{m}/2}\sum_{\alpha=1}^{N_\xi-1} c^{(\abs\gamma,\xi)}_\alpha\, B^{(\xi)}_\alpha(\xi)
\end{equation}
and
\begin{equation}
 \labeq{ydef}
  Y^{\abs\gamma}_n(\eta) \;=\; (1-\eta^2)^{\abs{m}/2}\sum_{\beta=1}^{N_\beta/2} c^{(\abs\gamma,\eta)}_\beta\, \tilde B^{(\abs\gamma,\eta)}_\beta(\eta)\,,
\end{equation}
\end{subequations}
with expansion coefficients $c^{(\abs\gamma,\xi/\eta)}_{\alpha/\beta}$.
Whereas the last B spline is excluded in \refeq{xdef} in order to fulfill zero- (or box-) boundary conditions, 
\begin{equation}
 \labeq{zerobc}
  X^{\abs\gamma}_n(\xi_\mrm{max}) \;=\; 0\qquad\forall\, n=1,\ldots,N_\xi-1\,,
\end{equation}
the $\eta$ basis \refeq{ydef} directly incorporates the parity (anti-)symmetry under the inversion transformation by choosing
\begin{equation}
 \tilde B^{(\abs\gamma,\eta)}_\beta(\eta) \;=\; B^{(\eta)}_\beta(\eta) + (-1)^{\abs m}\wp\,B^{(\eta)}_{N_\eta+1-\beta}(\eta)\,.
\end{equation}
Using this basis, its respective matrix elements w.\,r.\,t.\  \refeq{sham} are computed and the eigenfunctions \refeq{orb} along with their energies $\epsilon^\gamma_n(R)$ are determined by numerically solving a generalized eigenvalue problem that results from the Galerkin discretization of \refeq{oese}.
The results are then subsequently employed for forming the configurations \refeq{ci} of the two-electron basis.

In order to solve the two-electron Schr\"odinger equation
\begin{equation}
  \labeq{tese}
  \op{H}^\mrm{\left(H_2\right)}(R)\, \psi^{(\Gamma)}(\mbf r_1,\mbf r_2;R) \;=\; \epsilon^{(\Gamma)}_\psi(R)\,\psi^{(\Gamma)}(\mbf r_1,\mbf r_2;R)
\end{equation}
with
\begin{align}
 \begin{split}
  \op{H}^\mrm{\left(H_2\right)}(R) \;=\; &\op{H}^\mrm{\left(H_2{}^+\right)}(R\hat{\mbf e}_z) \otimes \op 1 \,+\, \op 1\otimes \op{H}^\mrm{\left(H_2{}^+\right)}(R\hat{\mbf e}_z)\,+\\
  &+\, \frac{1}{\abs*{\op{\mbf r}_1-\op{\mbf r}_2}}\,,
 \end{split}
\end{align}
the Coulomb repulsion term is expressed in terms of the von Neumann expansion~\cite{gosd:mehl69}.
The corresponding two-electron matrix elements
\begin{equation}
 \labeq{matelem}
 H^{(\Gamma)}_{n\bar n,n',\bar n'} \;=\; \delta_{nn'}\epsilon^\gamma_{n}+\delta_{\bar n\bar n'}\epsilon^{\bar\gamma}_{\bar n}+\Braket{\Phi^{(\Gamma)}_{n\bar n}|\frac{1}{\abs*{\op{\mbf r}_1-\op{\mbf r}_2}}|\Phi^{(\Gamma)}_{n'\bar{n}'}}
\end{equation}
are computed by Gaussian quadrature using a cache-optimized and parallelized implementation of the Ruedenberg approach~\cite{gosd:rued56} employing OpenMP. 
This allowed for the computation of matrix elements for several tens of thousands of configurations on the scale of a few hours on a 2 GHz, 48-core machine.

Once the matrix elements are determined, the Galerkin condition transforms \refeq{tese} into an eigenvalue equation
\begin{equation}
  \labeq{diag}
  \sum_{n'\bar n'} H^{(\Gamma)}_{n\bar n,n'\bar n'}(R)\, C^{(\Gamma)}_{n'\bar n'}(R) \;=\; \epsilon_\psi^{(\Gamma)}(R)\, C^{(\Gamma)}_{n\bar n}(R)\,,
\end{equation}
which can be numerically solved for each value of $R$ separately using the \texttt{DSYEVD} LAPACK routine.

The obtained solutions $\left(\epsilon_\psi^{(\Gamma)}, \left\{C^{(\Gamma)}_{n\bar n}\right\}\right)$ correspond to either discretized bound or continuum states.
Those bound states that have completely decayed at a chosen quasi-radial box size $\xi_\mrm{max}$ are well-represented by this approach and can be faithfully obtained and systematically improved by increasing $\xi_\mrm{max}$.
In contrast, the continuum states generally do not resemble the desired form that corresponds to the asymptotic behavior \refeq{mc_asymp}~\cite{gosd:bach01}.
The boundary conditions \refeq{zerobc} force the numerically obtained scattering states to form a more or less arbitrary linear combination of all degenerate channel contributions instead of giving the proper $S$-matrix basis scattering states $\psich$ for a definite channel $a$.
Furthermore, as only one solution per energy (corresponding to a certain combination of channels) is provided, it is generally not possible to straightforwardly retrieve the other channels at the \emph{same} energy~\cite{gosd:cort94, gosd:bach01}.
Thus, while still being useful for integrated spectra, cross sections differential in energy are cumbersome to get using this method, particularly when multiple, equally dominant channels are coupled.

\subsection{Free-boundary method}\labsec{fbm}

The FB method~\cite{gosd:lamb98, gosd:bros92, gosd:bros92a} remedies the short-comings laid out in \refsec{bound}.
It forces the numerical solution not to fulfill the single zero-boundary condition \refeq{zerobc} but instead to obey a set of $n_c$ linear-independently chosen boundary conditions that resemble all channels open at a \emph{freely} chosen value of energy $E$.
This method then allows for finding all $n_c$ linear combinations of partial-wave solutions which subsequently can be fitted at a sufficiently large value of $\xi_\mrm{max}$ to the real-valued $K$-matrix (or standing-wave) boundary condition
\begin{align}
\begin{split}
  \labeq{mc_kasymp}
  \psi&^{(\Gamma,s)}_{E',\,a\lambda}(\mbf r_1,\mbf r_2;R) \underset{\xi_1\to\infty}{\sim} \op{\mathcal A}^{(S)} \sum\limits_{b=1}^{n_c} \sum_{\lambda'=|\bar m_b|}^\infty \frac{1}{R\, \xi_1 \sqrt{p'_b}}\, \times\\
  &\times\, \phi^{\gamma_b}_{n_b}(\mbf r_2;R)\,\Upsilon^{\bar m_b}_{\lambda'}(\eta_1,\varphi_1\,;\, C_b')\,\times\\
  &\times\, \bigg(\delta_{ab}\delta_{\lambda\lambda'}\,\sin\theta^{|\bar m_b|}_{b\lambda'} \,+ \,K^{(\Gamma)}_{a\lambda,b\lambda'}\,\cos\theta^{|\bar m_b|}_{b\lambda'}\bigg)\,,
\end{split}
\end{align}
where the $n_c\times n_c$-dimensional $K$ matrix is related to the $S$ matrix by $\mbf S = (\mbf 1 - \im\mbf K)^{-1}(1 + \im\mbf K)$.
By imposing the boundary condition \refeq{mc_kasymp} already at finite values $\xi_1=\xi_\mrm{max}$, the electron-repulsion interaction is effectively truncated beyond this value, which is the predominant meaning of the artificial parameter $\xi_\mrm{max}$ in the context of the FB method.
This is in contrast to the zero-boundary condition approach, where this value typically describes the onset of infinitely high walls, hence discretizing the continuum.

More concisely, the FB method can be summarized as follows.
The homogeneous scattering problem 
\begin{subequations}
\labeq{inhomo}
\begin{equation}
  \left(\op H^\mrm{\left(H_2\right)} - E\right)\,\tilde\psi^{(\Gamma)}_{E',\,a\lambda}(\mbf r_1, \mbf r_2) \;\equiv\; \op A(E)\,\tilde\psi^{(\Gamma)}_{E',\,a\lambda} \;=\; 0
\end{equation}
with inhomogeneous boundary condition
\begin{equation}
  \tilde\psi^{(\Gamma)}_{E',a\lambda}(\mbf r_1, \mbf r_2)\Big|_{\xi_1=\xi_\mrm{max}} \;\overset{!}{=}\; \op{\mathcal S}^{(\Gamma)}\left[\phi^{\gamma_a}_{n_a}(\mbf r_2)\,f^{\bar\gamma_a}_a(\mbf r_1)\right]\Big|_{\xi_1=\xi_\mrm{max}} 
\end{equation}
\end{subequations}
and freely chosen values of $E>E_\mrm{thresh}$ (with $E_\mrm{thresh}$ being the lowest-lying threshold) is easily seen to be equivalent to an inhomogeneous problem 
\begin{subequations}
   \labeq{homo}
  \begin{equation}
   \op A(E)\, \tilde\psi^{(\Gamma)'}_{E',\, a\lambda}(\mbf r_1, \mbf r_2) \;=\; -\op A(E)\, \op{\mathcal S}^{(\Gamma)}\left[\phi^{\gamma_a}_{n_a}(\mbf r_2)\,f^{\bar\gamma_a}_a(\mbf r_1)\right]
 \end{equation}
 with a homogeneous boundary condition given by
 \begin{equation}
   \tilde\psi^{(\Gamma)'}_{E',\,a\lambda}(\mbf r_1,\mbf r_2)\Big|_{\xi_{1/2}=\xi_\mrm{max}} \;\overset{!}{=}\; 0\,,
 \end{equation}
\end{subequations}
where the molecular symmetry adaptor $\hat{\mathcal S}^{(\Gamma)}$, \refeq{adapter}, has been used.
The respective solutions of Eqs. (\ref{eq:inhomo}) and (\ref{eq:homo}) are related by
\begin{align}
 \begin{split}
  \tilde\psi^{(\Gamma)}_{E',a\lambda}(\mbf r_1,\mbf r_2) \;=\; &\tilde\psi^{(\Gamma)'}_{E',a\lambda}(\mbf r_1,\mbf r_2) \,+\\
  &+\, \op{\mathcal S}^{(\Gamma)}\left[\phi^{\gamma_a}_{n_a}(\mbf r_2)\,f^{\bar\gamma_a}_a(\mbf r_1)\right]\,,
 \end{split}
\end{align}
for some arbitrary (non-zero) boundary function, which can be conveniently chosen to feature the last B spline omitted in \refeq{xdef} for the quasi-radial and a spheroidal harmonic, \emph{cf.}\ \refeq{sph_harm}, for the quasi-angular part,
\begin{equation}
 \labeq{boundf}
  f^{\bar\gamma_a}_a(\xi,\eta,\varphi) \;=\; \Upsilon^{\bar m}_{\bar\lambda}(\eta,\varphi;C_a')\, B^{(\xi)}_{N_\xi}(\xi)\,.
\end{equation}
Note, the implicit energy dependency of $f^{\bar\gamma_a}_a(\xi,\eta,\varphi)$ introduced by the parameter $C_a'$ in the spheroidal harmonic is not a requirement of the method \emph{per se}.
In fact, it is only necessary for the boundary functions \refeq{boundf} to be of proper molecular symmetry defined by the channel and to be linearly independent from each other, the latter being the case for any specific value of energy.
Thus, from a numerical standpoint, it is advisory to keep the energy fixed to any convenient choice in order to prevent an expensive re-computation of the integrals in \refeq{qvec} below for each value of $E$.

Employing the same finite CI expansion \refeq{ci} and imposing the Galerkin condition on the residual yields the linear problem
\begin{equation}
  \labeq{fbm}
  \sum_{n'\bar n'} \left(H^{(\Gamma)}_{n\bar n,n'\bar n'} \,-\,  \delta_{nn'}\delta_{\bar n\bar n'}\, E\right) C^{(\Gamma)}_{n'\bar n'}(E) \;=\; q^{(\Gamma)}_{n\bar n}(E)
\end{equation}
with $H^\Gamma_{n\bar n,n'\bar n'}$ being a component in the block of the Hamilton matrix corresponding to molecular symmetry $\Gamma$ and boundary-vector elements
\begin{equation}
  \labeq{qvec}
  q^{(\Gamma)}_{n\bar n}(E) \;=\; \Braket{\Phi^{(\Gamma)}_{n\bar n}|\left(E \,-\, \op{H}^\mrm{(H_2)}\right)\,\op{\mathcal S}^{(\Gamma)}|\phi^{\gamma_a}_{n_a}f^{\bar\gamma_a}_a}\,,
\end{equation}
which are the only additional integrals that need to be computed for the FB method.
While the specific choice of $B^{(\xi)}_{N_\xi}$ in \refeq{boundf} is by no means essential, it is, given the existing codes, convenient and involves the computation of only few matrix elements for those B splines with overlapping support.

In the end, the symmetric diagonalization problem \refeq{diag} for the zero-boundary conditions \refeq{zerobc} is replaced by a symmetric-indefinite linear problem \refeq{fbm} obeying inhomogeneous boundary conditions \refeq{inhomo} while employing the same matrix elements \refeq{matelem} and, therefore, most of the original code.

As was pointed out in Ref.~\cite{gosd:bros92a}, it is worthwhile to transform the symmetric linear problem \refeq{fbm} into a positive-definite one by matrix multiplication of both sides of the equation by $\left(\mbf H^\Gamma - E\, \mbf 1\right)$.
If the linear problem has to be solved for many values of $E$, it is typically most efficient to pre-compute and store the energy-independent matrix $\left(\mbf H^\Gamma\right)^2$ once and subsequently add together the mixed terms for each value of energy.
Computing the matrix product is a numerically expensive operation, which has been accomplished utilizing the \texttt{DSYRK} LAPACK routine~\cite{gosd:lapa}.
Afterwards, the system can be solved by performing a regular Cholesky decomposition, which needs to be computed only once per energy, \emph{e.g.}, by using the \texttt{DPOTRF} LAPACK routine~\cite{gosd:lapa}, but is reused for all degenerate channel solutions.
Alternatively, if the total number of points on the energy-loss grid is sufficiently large relative to the matrix dimension of the system \refeq{fbm}, it is more time-efficient to compute an exact diagonalization of $\mbf H^\Gamma$, \emph{e.\,g.}, by the \texttt{DSYEVD} LAPACK routine~\cite{gosd:lapa} once and subsequently solve the linear problem in the eigenbasis, hence reducing the numerical cost per energy value from $O(N^3)$ to $O(N^2)$ for an $N\times N$-dimensional problem.

Once a set of $n_c$ solutions $\left\{\tilde\psi^{(\Gamma)}_{E',a\lambda}\right\}$ has been obtained from \refeq{fbm} for a certain energy $E'$, they can be projected on the channels, evaluated, and used for a two-point fit near $\xi_1=\xi_\mrm{max}$ to
\begin{align}
 \begin{split}
  \labeq{num_asymp}
  \tilde\psi&^{(\Gamma)}_{E',\,a\lambda}(\mbf r_1,\mbf r_2;R) \underset{\xi_1\to\infty}{\sim} \op{\mathcal A}^{(S)} \sum\limits_{b=1}^{n_c} \sum_{\lambda'=|\bar m_b|}^\infty \frac{1}{R\, \xi_1 \sqrt{p'_b}}\, \times\\
  &\times\, \phi^{\gamma_b}_{n_b}(\mbf r_2;R)\,\Upsilon^{\bar m_b}_{\lambda'}(\eta_1,\varphi_1\,;\, C_b')\,\times\\
  &\times\, \bigg(A^{(\Gamma)}_{a\lambda,b\lambda'} \,\sin\theta^{|\bar m_b|}_{b\lambda'} \,+ \, B^{(\Gamma)}_{a\lambda,b\lambda'}\,\cos\theta^{|\bar m_b|}_{b\lambda'}\bigg)\,,
 \end{split}
\end{align}
(with $\theta_b$ being the same as in \refeq{sph_phase}).
This is repeated for each open channel $b\lambda'$ in order to determine the fitting coefficients $A_{a\lambda,b\lambda'}$ and $B_{a\lambda,b\lambda'}$, which relate to the truncated $K$ matrix by $\mbf K = \mbf A^{-1}\mbf B$, that is, the standing-wave solution with boundary condition \refeq{mc_kasymp} is
\begin{align}
 \begin{split}
  \psi&^{(\Gamma, s)}_{E',\, a\lambda}(\mbf r_1,\mbf r_2;R) \\
  &=\; \sum_{b=1}^{n_c}\sum\limits_{\lambda'=|\bar m_b|}^\infty \left(\mbf A^{-1}\right)_{a\lambda,b\lambda'}\, \tilde\psi^{(\Gamma)}_{E',\, b\lambda'}(\mbf r_1,\mbf r_2;R)\,.
 \end{split}
\end{align}
In order to further increase the accuracy of the fitting procedure, particularly for energies close to the threshold, the $\sin$ and $\cos$ functions may be replaced by the single-center, (ir-)regular Coulomb wave functions $F_{m+\lambda}(1/p'_b, p_b' R\sqrt{\xi^2-1}/2)$ (and correspondingly for $G_\ell$), or, more accurately, by the proper two-center prolate spheroidal Coulomb wave functions, if available~\cite{gosd:sing23, gosd:volk26}.
Particularly, in the case of the near-threshold cross sections, using any of these asymptotic functions allows for significantly smaller values of $\xi_\mrm{max}$. 

\subsection{Form-factor matrix elements}\labsec{form}

As is apparent from Eqs. (\ref{eq:ci}), (\ref{eq:conf1}), (\ref{eq:conf2}), and (\ref{eq:orb}), the integrals in \refeq{avg_cross_sec} conveniently separate into sums and products of one-dimensional integrals, which can be efficiently computed even for a large number of configurations.
The radial and angular spheroidal wavefunctions required for \refeq{expterm} are numerically evaluated using the \texttt{profcn} routine~\footnote{The code is made publicly available at \url{https://github.com/MathieuandSpheroidalWaveFunctions/prolate_swf}.}~\cite{gosd:bure02, gosd:bure04}, which provides excellent accuracy for a large range of values of $L$, $M$, and $K$.

In this work, ionic orbitals from the same basis set are used to form the CI series representing both the initial and the scattering states. 
For a single-particle operator, the number of integrals to be computed is therefore restricted to those configurations $\Phi^\Gamma_{n\bar n}$ and $\Phi^{\Gamma'}_{n'\bar n'}$ for which any of the rules
\begin{equation}
 \labeq{cases}
  n=n',\ n=\bar n',\ \bar n=n',\ \text{and } \bar n=\bar n'
\end{equation}
applies.
Furthermore, the individual terms of the expansion \refeq{prosphexp} feature selection rules that further restrict the number of matrix elements between certain symmetries.
First of all, for $\bar n=\bar n'$, the $\varphi$ integrations are only non-vanishing if
\begin{equation}
 \labeq{mselect}
  M\,+\,m\,+\,m'\;=\; 0
\end{equation}
(and analogously for the other cases in \refeq{cases}).
Furthermore, one finds for the integrals w.\,r.\,t.\  the $\eta$ coordinate
\begin{equation}
  \int\limits_{-1}^1 B^{(\abs\gamma,\eta)}_\beta(\eta) \Upsilon^M_L(\eta,\varphi;C) B^{(\abs{\gamma'},\eta)}_{\beta'}(\eta)\, \eta^2\,\rmd\eta \;=\; 0\,,
\end{equation}
if the integrand is an odd function in $\eta$.
The $\eta^2$ factor may also be replaced by a factor of $1$ (having the same symmetry), which both originate from the prolate-spheroidal volume element.
Therefore, if $L+M$ is even (odd), the integral vanishes if the product $B^{(\abs\gamma,\eta)}_\beta(\eta) B^{(\abs{\gamma'},\eta)}_{\beta'}(\eta)$ is odd (even).
In turn, a single function $B^{(\abs\gamma,\eta)}_\beta(\eta)$ is even if $m$ is even and $\wp$ is \emph{gerade} or if $m$ is odd and $\wp$ is \emph{ungerade}; otherwise, it is odd.

Once the one-dimensional integrations are performed by using Gaussian quadrature, they are combined into the transition matrix elements \refeq{avg_cross_sec}.
Due to the selection rule \refeq{mselect}, only scattering states of total angular momentum up to $\Lambda=M_\mrm{max}$ have to be computed, where $M_\mrm{max}$ is a numerical cutoff value for the expansion \refeq{prosphexp}.
A satisfactory value for $M_\mrm{max}$ depends particularly on the magnitude of the momentum-transfer modulus $K$.
For the largest considered value of $K=8.0$ $a_0{}^{-1}$, calculations using $\Lambda_\mrm{max}=19$ have been employed.
No comparable selection rule exists for $L$, for which the cutoff value $L_\mrm{max}=\Lambda + q_{\eta,\, \mrm{max}}$ has been found to be adequately converged, where $q_{\eta,\, \mrm{max}}=20$ refers to the largest number of $\eta$ nodes present within the configuration series.

Instead of computing the (complex-valued) transition amplitudes \refeq{ampl} w.\,r.\,t.\  the solution given in the $S$-matrix basis, it is more convenient to employ the real-valued solutions $\tilde\psi^{(\Gamma)}_{a \epsilon_a'}(\mbf r_1,\mbf r_2;R)$ obtained from \refeq{fbm} with asymptotic behavior \refeq{num_asymp} for computing the amplitudes $\tilde A^{M}_{L,\,a\lambda m}$.
Using the fitting coefficients $A_{ab}$ and $B_{ab}$ from \refsec{fbm}, the proper amplitudes are then found to be
\begin{align}
 \begin{split}
  A&^{M(-)}_{L,\,a\lambda m} \\
  &=\; \sum_{b=1}^{n_c}\sum_{\lambda'=|\bar m_b|}^\infty\left(\left[\mathbf A \,+\, \im\, \mathbf B\right]^{-1}\right)_{a\lambda,b\lambda'} \tilde A^{M(-)}_{L,\,b\lambda' \bar m_b}\,.
 \end{split}
\end{align}

Special care must be taken for $\Gamma\to\Gamma$ transitions around $K\approx 0$, for which $\exp(\im\, \mbf K\cdot\mbf r_i) \to 1 \,+\, O(K)$.
For inelastic transitions, \emph{e.g.}, from a bound state to the continuum, the overlap of the involved states should vanish by the orthogonality inherent to the solutions of the Schr\"odinger equation.
However, if the bound and scattering states are constructed using different CI series \refeq{ci} (which typically is expedient in view of the quite differing nature of these states), their respective truncated Hilbert spaces generally are not perfectly overlapping and the orthogonality is only approximate.
As such, residual spurious contributions of order $O(1)$ are \enquote{overshadowing} the $O(K)$ contributions, which significantly affects the overall transition amplitudes for small values of $K$.
A rather pragmatic workaround is to explicitly subtract the zero-order contribution from the exponential, which hence directly cancels the noise and has been found to work seamlessly.

\subsection{Numerical parameters}

For the computations a quasi-radial cutoff of $\xi_\mrm{max}=50\, a_0$ has been found to be a good compromise for representing the localized ground state and the oscillatory high-energy continuum simultaneously.
For the B-spline basis, a total of $N_\xi-1=200$ uniformly distributed functions of order 7 for $B^{(\xi)}_\alpha$ and $N_\eta/2 = 25$ ones of order 5 for $B^{(\eta)}_\beta$ are employed.
The ground-state configuration series are summarized in \reftbl{gsci}, which, excluding over-counting due to double occurrences, gives rise to a total number of 90,600 configurations.
The series is not optimized in the sense of featuring only the most relevant configurations (in terms of impact on correlation energy), but was heuristically determined from a full-CI computation using a significantly smaller box, which allowed for determining those symmetries which contain the largest population.
Including up to ten terms in the von Neumann expansion~\cite{gosd:mehl69,gosd:vann04} of the electron-repulsion operator results in a ground-state energy of about $-1.17407$ a.u., to be compared to the essentially exact reference value of $-1.17447$ a.u.~\cite{gosd:kolo65}.
Thus, about 99\,\% of the electronic correlation energy is reproduced.

Concerning the scattering wavefunctions, each ionic channel $\phi^{\gamma}_a$ featured in the computation is represented by a series of configurations with one electron fixed to the ionic-channel state and the other electron taking any of the $N_\xi-1=200$ available ionic states.
Furthermore, for each of the ionic channels, there are in total $N_\eta=20$ possible quasi-angular channels.
Series of the $8\times 8$ energetically lowest-lying bound states are included in the basis as well in order to approximate the doubly-excited bound-state character of the continuum wavefunction for energies in the vicinity of the doubly excited autoionizing states.
The $n_c$ channel functions $\phi^{\gamma_a}_{n_a}$ are chosen to be the energetically lowest states for the given value of internuclear separation $R=1.4$ $a_0$ where the 14 ionic channels 1s$\sigma_\mrm{g}$, 2p$\sigma_\mrm{u}$, $\pm$2p$\pi_\mrm{u}$, 2s$\sigma_\mrm{g}$, 3p$\sigma_\mrm{u}$, 3d$\sigma_\mrm{g}$, $\pm$3d$\pi_\mrm{g}$, $\pm$3d$\delta_\mrm{g}$, $\pm$3p$\pi_\mrm{u}$, and 3s$\sigma_\mrm{g}$ below the 10th threshold are considered in the FB computations.
Furthermore, for each of these 14 ionic channels, the lowest 10 pseudo-angular channels (with 0, 2, \ldots, 18 nodes for even and 1, 3, \ldots, 19 for odd $\eta$ wavefunctions $Y^{\abs\gamma}_n(\eta)$ \refeq{ydef}, respectively) have been included, resulting in a total number of 140 channels, each being represented by all of their respective ionic orbitals.
Finally, in order to properly describe the non-zero boundary condition, the remaining highest 12 pseudo-angular channels of each ionic channel also contain series for the 8 energetically \emph{highest} ionic orbitals, resulting in a total amount of between 22,610 configurations for $^1\Sigma^+_\mrm{g}$ symmetry and up to 34,162 for the symmetries with $\Lambda\ge 4$.
All computations done in this work are restricted to the fixed-nuclei approximation at equilibrium internuclear-separation $R=1.4$ $a_0$, which results in the neglect of the vibronic degrees of freedom in \refeq{ampl}.
A further investigation which features the effects of nuclear motion more accurately is planned for the future and thus not subject of this work.
Additionally, the code can also be used for a description of the limit $R\to 0$, providing the data for helium atoms.
The results are very promising and are planned to be published in a separate work as well.

\section{Results}\labsec{results}

\begin{table}[t]
\begin{tabular}{llll}
  \toprule 
  $e^-_1$ basis & \hspace{1.0cm}$\otimes$\hspace{0.5cm} & $e^-_2$ basis & \\
  \midrule 
  series & sym. & series & sym. \\
  \midrule 
   1 --  200 & s$\sigma_\mrm{g}$ & 1  --  200 & s$\sigma_\mrm{g}$ \\
   1 --  200 & p$\sigma_\mrm{u}$ & 1 --  200 & p$\sigma_\mrm{u}$ \\
   1 --  \phantom{1}50 & s$\sigma_\mrm{g}$ & 1  --  100 & d$\sigma_\mrm{g}$ \\
   1 --  \phantom{1}50 & d$\sigma_\mrm{g}$ & 1  --  100 & d$\sigma_\mrm{g}$ \\
   1 --  \phantom{1}50 & f$\sigma_\mrm{u}$ & 1  --  100 & f$\sigma_\mrm{u}$ \\
   1 --  \phantom{1}50 & g$\sigma_\mrm{g}$ & 1  --  \phantom{1}50 & g$\sigma_\mrm{g}$ \\
   \midrule
   1 --  200 & p$\pi_\mrm{u}$ & 1 --  200 & p$\pi_\mrm{u}$ \\
   1 --  100 & d$\pi_\mrm{g}$ & 1 --  100 & d$\pi_\mrm{g}$ \\
   1 --  \phantom{1}50 & f$\pi_\mrm{u}$ & 1 -- 100 & f$\pi_\mrm{u}$ \\
   1 --  \phantom{1}50 & g$\pi_\mrm{g}$ & 1 --  \phantom{1}50 & g$\pi_\mrm{g}$ \\
   \midrule
   1 --  \phantom{1}50 & d$\delta_\mrm{g}$ & 1 -- \phantom{1}50 & d$\delta_\mrm{g}$ \\
   1 --  \phantom{1}50 & f$\delta_\mrm{u}$ & 1 -- \phantom{1}50 & f$\delta_\mrm{u}$ \\
   1 --  \phantom{1}50 & g$\delta_\mrm{g}$ & 1 -- \phantom{1}50 & g$\delta_\mrm{g}$ \\
   \midrule
   1 --  \phantom{1}50 & f$\phi_\mrm{u}$ & 1 -- \phantom{1}50 & f$\phi_\mrm{u}$ \\
   1 --  \phantom{1}50 & g$\phi_\mrm{g}$ & 1 -- \phantom{1}50 & g$\phi_\mrm{g}$ \\
   \bottomrule
\end{tabular}
\caption{Two-electron CI configuration series used for the ground states in the FB computations. The numbers indicate the index of the ionic single-particle orbitals ordered w.\,r.\,t.\  increasing energy including both bound and box-discretized continuum states.}
\labtbl{gsci}
\end{table}

\begin{figure*}[t]
  \subfigure[][]{
  \labfig{gosd_1}
  \includegraphics[width=0.49\textwidth]{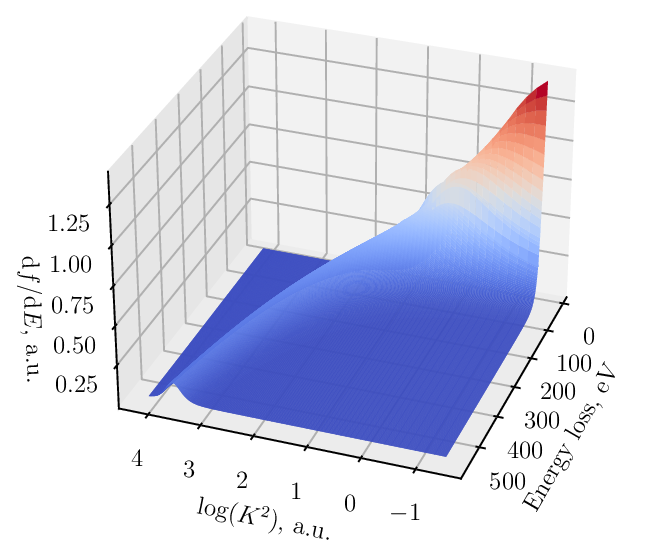}}
  \subfigure[][]{
  \labfig{gosd_2}
  \includegraphics[width=0.49\textwidth]{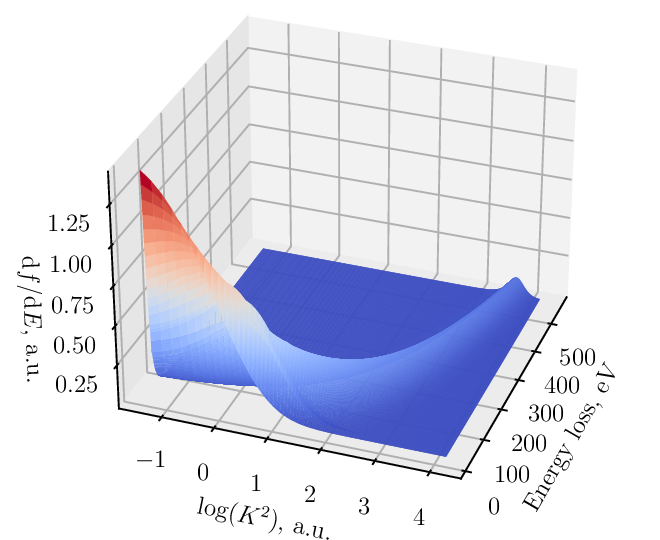}}
  \caption{The Bethe surface computed using Eqs. (\ref{eq:avg_cross_sec}) and (\ref{eq:egosd}) for randomly oriented H$_2$ molecules with nuclei fixed at  internuclear distance $R=1.4$ $a_0$. In order to show both sides of the \enquote{Bethe ridge}, the axes for momentum transfer and energy loss in \subref{fig:gosd_2} are inverted w.\,r.\,t.\  \subref{fig:gosd_1}.
  The color-coding for the values of the GOSD is redundant with the reading on the vertical axis and has just been given for clearer visibility, running from dark blue (0 a.u.) to dark red (1.53 a.u.).}
  \labfig{bethe}
\end{figure*}

\begin{figure}
  \subfigure[][]{
  \labfig{mconv}
  \includegraphics[width=0.23\textwidth]{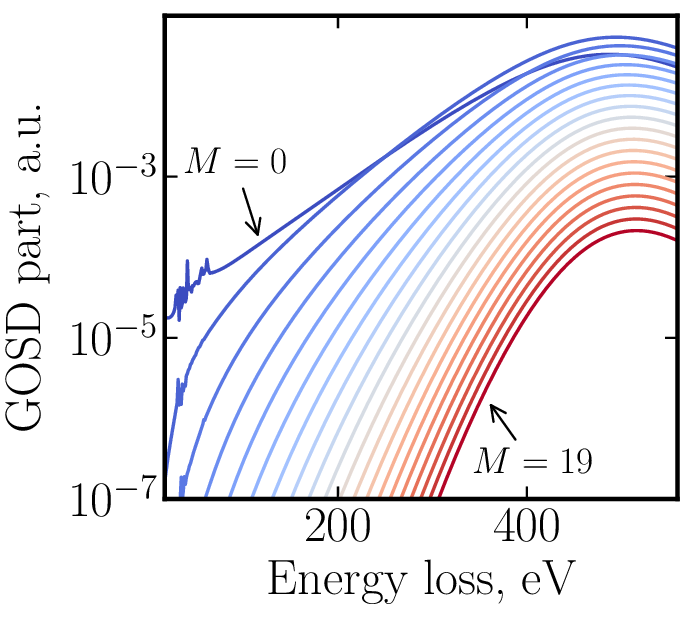}}
  \subfigure[][]{
  \labfig{lconv}
  \includegraphics[width=0.23\textwidth]{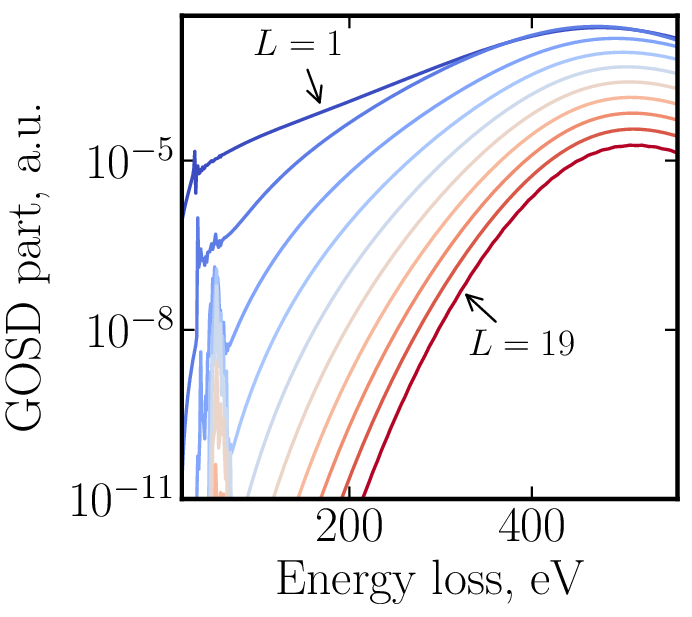}}
  \subfigure[][]{
  \includegraphics[width=0.47\textwidth]{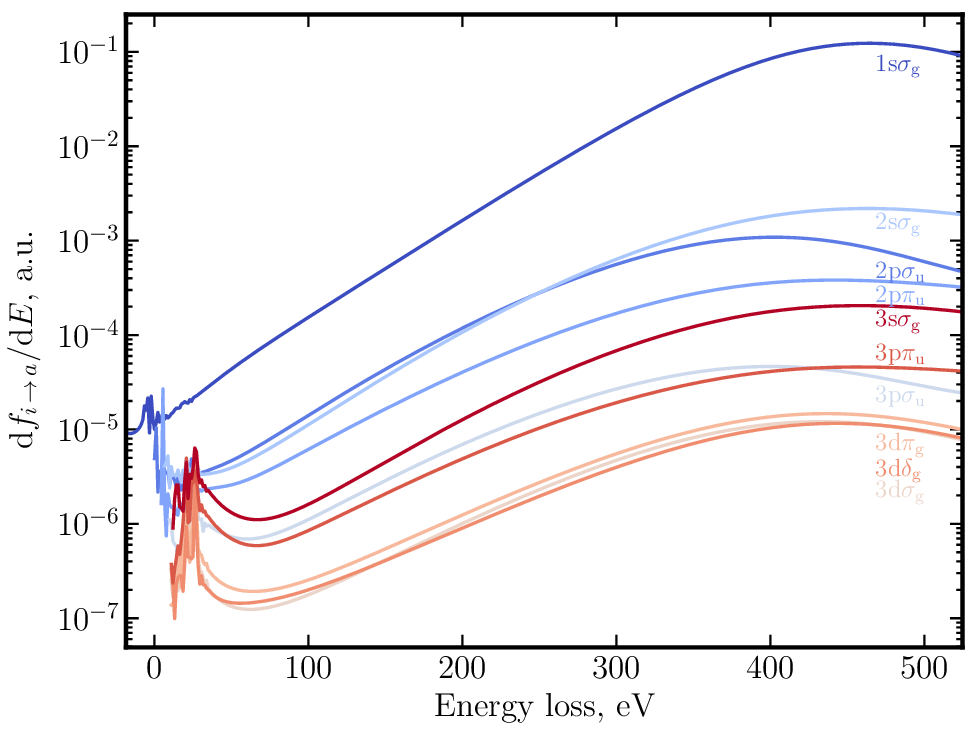}
  \labfig{chan_cont}
  }
  \caption{Partial summations of \refeq{avg_cross_sec} in terms of the GOSD at $K=6.0$ $a_0{}^{-1}$. In \subref{fig:mconv}, the contributions of all $\Lambda_\mrm{max}+1=20$ terms are shown, the color spectrum going from dark red for $\Lambda=0$ to dark blue for $\Lambda=19$.
  The plot in \subref{fig:lconv} depicts the eight different $q_{\eta,\,\mrm{max}}=1,3,5,\ldots,19$ $L$ contributions to the ${}^1\Sigma^+_\mrm{u}$ symmetry, corresponding to the quasi-angular channels.
  The color spectrum goes from dark red (1 $\eta$ node) to dark blue (19 $\eta$ nodes).
  In turn, \subref{fig:chan_cont} shows the channel-resolved partial GOSD $\frac{\rmd f_{i\, \to\, a}}{\rmd E}$.
  The different lines are labelled by the H$_2{}^+$ ionic channel.}
  \labfig{convergence}
\end{figure}

\begin{figure}[t]
  \includegraphics[width=0.48\textwidth]{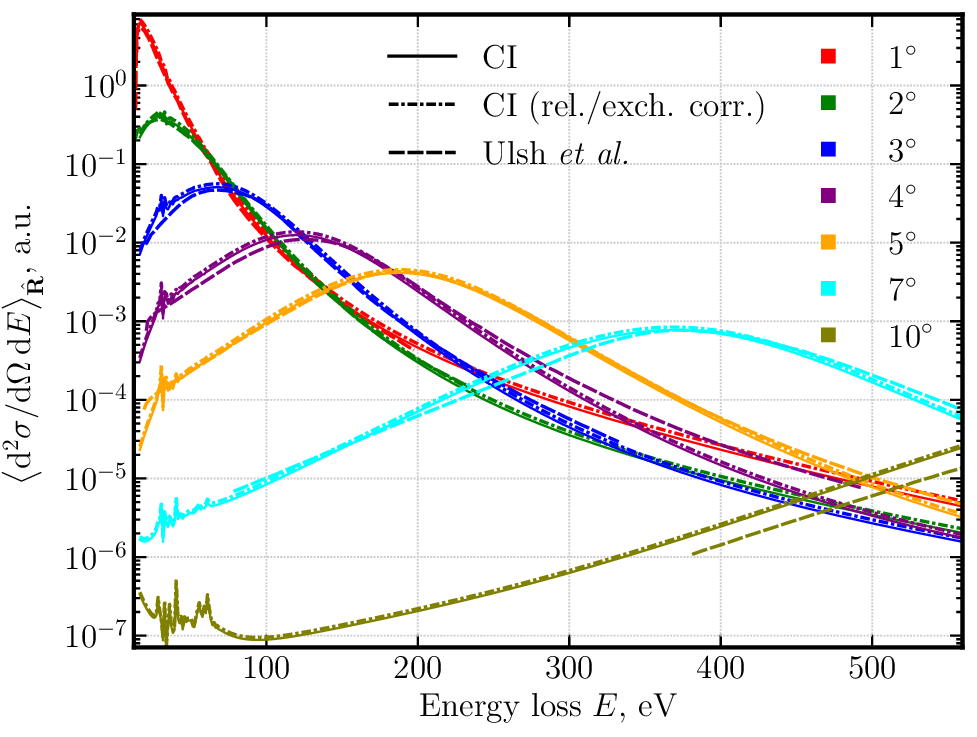}
  \caption{Comparison of the orientation-averaged doubly-differential cross section of electron impact to the reproduced experimental data of Ulsh \emph{et al.}~\cite{gosd:ulsh74}
  The solid lines show the CI result of the present work using \refeq{egosd}, with different colors for varying values of scattering angle $\vartheta_p$ (\emph{cf.}\ \refeq{angle}) as indicated in the legend.
  Dash-dotted lines indicate the results including relativistic and exchange corrections.
  The dashed curves (of same color) depict the experimental data.}
  \labfig{bonham}
\end{figure}

\subsection{Bethe surface}

In \reffig{bethe}, the obtained Bethe-surface plots for the covered parameter space up to $K=8.0$ $a_0{}^{-1}$ and $E\approx 558$ eV are given, with data points sampled on a 500 $\times$ 150-sized grid for values of $E$ and $K$, respectively~\footnote{The GOSD data presented in \reffig{bethe} is made publicly accessible at \url{https://doi.org/10.5281/zenodo.22808143}.}.
It shows the expected three regimes, namely the ionization-potential dominated low-energy region driven by dipole-allowed transitions, the high-energy binary-encounter region driven by several beyond-dipole transitions giving rise to the \enquote{Bethe ridge}~\cite{gosd:inok71} following approximately $E = K^2/2$, and an intermediate region, where dipole allowed and forbidden transitions compete~\cite{gosd:lahm88}.
In particular, for $K\to 0$, it is also possible to extract values closely related to the single-photon ionization cross-section~\cite{gosd:froe85, gosd:volk26a}.

\begin{figure*}[t]
  \subfigure[][]{
  \labfig{zurales_high}
  \includegraphics[width=0.47\textwidth]{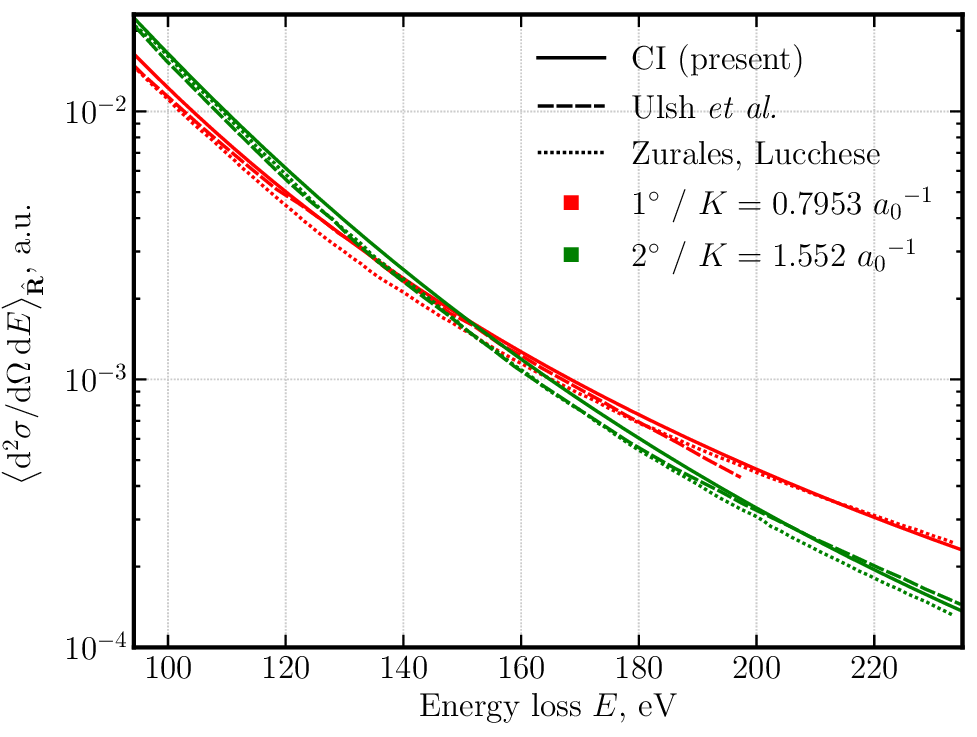}}
  \subfigure[][]{
  \labfig{zurales_low}
  \includegraphics[width=0.47\textwidth]{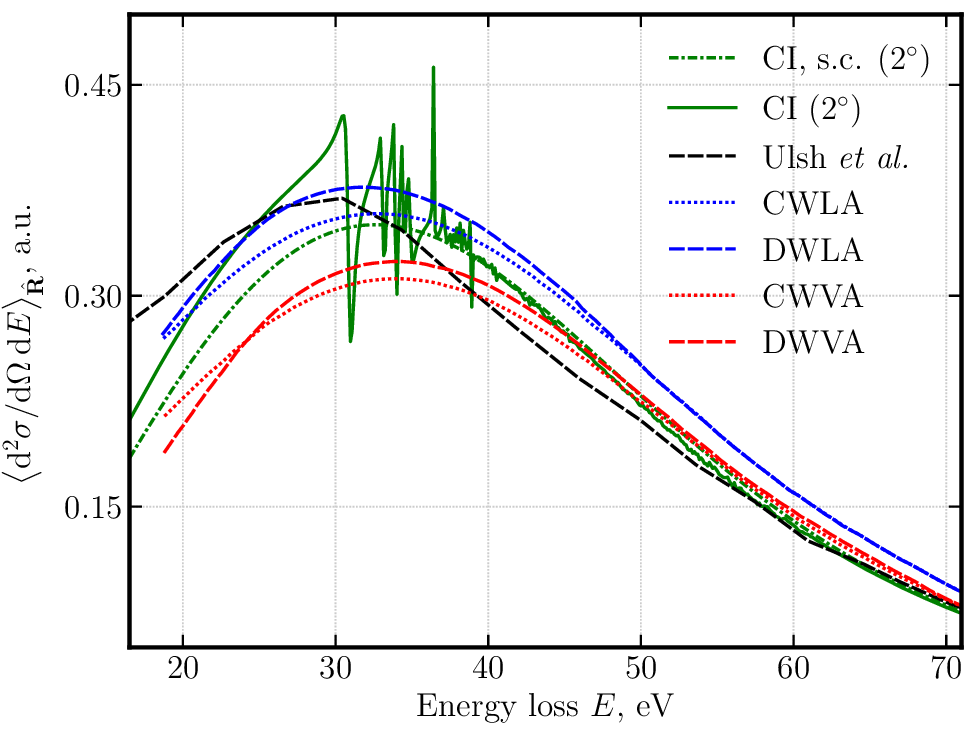}}
  \caption{Same as \reffig{bonham}, but additionally showing the comparison to the theoretical data of Ref.~\cite{gosd:zura88}
  The solid lines stand for the results of this work, the dashed lines are the reproduced experimental results of Ref.~\cite{gosd:ulsh74}
  In \subref{fig:zurales_high}, the dotted line shows the DWVA result of Ref.~\cite{gosd:zura88} for the two different values of $K$ closest to the scattering angles under consideration.
  The red and green colors indicate the respective scattering angle or momentum transfer as in \reffig{bonham}.
  The plot in \subref{fig:zurales_low} compares the distorted and Coulomb-wave results for length and velocity forms from Ref.~\cite{gosd:zura88} to the experimental~\cite{gosd:ulsh74} and CI data at lower values of energy on a ten times finer energy grid.
  Additionally, the CI result with reduced correlation employing only a single (ionic) channel (s.c.) is given by the dash-dotted line.}
  \labfig{zurales}
\end{figure*}

Both, the convergence properties as well as the individual contributions to the summation in \refeq{avg_cross_sec} for the DDCS at $K=6.0$ $a_0{}^{-1}$ are further exemplified in \reffig{convergence}.
It can be seen in \reffig{mconv} that, even deep within the binary-encounter regime, the dipole-allowed ${}^1\Pi_\mrm{u}$ symmetry still provides the largest contribution at the Bethe ridge, whereas the ${}^1\Sigma^+_\mathrm{g/u}$ contributions are being suppressed below the ${}^1\Phi_\mathrm{g/u}$ contributions at high energy loss.
As the $\Lambda=19$ contribution accounts for less than 1\% of the total Bethe ridge and, with the exception of $\Lambda=0$, the contributions are monotonically and seemingly exponentially decreasing with $\Lambda$, no error beyond 1\% is to be expected by truncating the summation at $\Lambda_\mrm{max}$.
Similar conclusions might be drawn from \reffig{lconv}.
Each $L$ term in \refeq{ampl} has its dominant contribution from the $q_\eta = L - m$ quasi-angular channel, which can be observed for all molecular symmetries alike and is shown here for the $^1\Sigma^+_\mrm{u}$ case.
The contribution of $L=18$  accounts for less than 1\% of that for $L=0$, providing an pessimistic upper bound for the truncation error below this value.
Therefore, inclusion of more $\eta$ channels is expected to improve the Bethe ridge by a similar order of magnitude.

The question of how the different partial ionic channels contribute to the total DDCS is further elucidated in \reffig{chan_cont}, which depicts the channel-resolved GOSD $\rmd f_{i\,\to\, a}/\rmd E$ corresponding to each ionic state whose associated channel opens at one of the ten thresholds considered in this work.
It can be seen that the ionization channel's threshold energy itself is not correlated with the overall contribution, the only exception being the ionic ground state $1s\sigma_\mrm{g}$, which has been found to be the most-dominant channel for all energy losses, and values of $K$ by around two orders of magnitude.
Remarkably, the Bethe-ridge position for the less dominant channels appears to vary and is generally found at smaller energy losses than for the dominant channel.
Although no precise reason has been found for the specific order of the channel contributions, it is, however, not entirely unpredictable.
For this purpose, it is instructive to examine the populations of the initial-state configuration.
In particular, by adding together the populations of those configurations which contain the ionic state associated with the channel, one finds for the eight most populated configurations 98.4, 4.85, 0.97, 0.47, 0.35, 0.15, 0.09, and 0.07\% of the total population to be contributed by configurations containing the 1s$\sigma_\mrm{g}$, 2s$\sigma_\mrm{g}$, 2p$\sigma_\mrm{u}$, 3s$\sigma_\mrm{g}$, 2p$\pi_\mrm{u}$, 4s$\sigma_\mrm{g}$, 3p$\sigma_\mrm{u}$, and 3p$\pi_\mrm{u}$ orbitals, respectively.
If one refrains from the absence of the $4s\sigma_\mrm{g}$ channel in the computation as well as the swapped order of the 3s$\sigma_\mrm{g}$ and 2p$\pi_\mrm{u}$ contributions, one sees that these numbers (which are entirely determined by the initial-state computation alone) appear to \enquote{echo} the significance of the channel contributions within the total GOSD to a certain degree.
From a numerical standpoint, such information might indeed be useful in order to find a better criterion for truncating the channels than their threshold energy.
Physically, one may interpret this resemblance as the result of kicking off an electron from any of the most significant ground state configurations, leaving behind an ion whose excited-state distribution may be inferred from the initial-state configurations and, potentially, \emph{vice-versa}.
This concept is, in fact, also closely related to the \enquote{spectroscopic factor} or \enquote{pole strength} quantities, as used in the context of EMS experiments~\cite{gosd:naka25}, which are, particularly for larger molecules, subject of ongoing research.
Notably, the order of the ionic states as seen in \reffig{chan_cont} can also be observed in the EMS outcomes of Lermer \emph{et al.}~\cite{gosd:lerm97}, which, however, used scattering energies at which the accuracy of the FBA is questionable~\cite{gosd:taka03}, though later theoretical investigations found no evidence of postulated second Born effects~\cite{gosd:capp06}.
In any case, as can be seen in \reffig{chan_cont}, no essential channel that could have been erroneously truncated appears to be missing.
Hence, convergence on the sub-percent level for the total GOSD may be assumed for the present results.

\subsection{Experimental and theoretical verification}

In order to estimate the agreement to experimental results, the GOSD is further compared to the data provided by Ulsh \emph{et al.}~\cite{gosd:ulsh74} for electrons incident with 25 keV, as presented in \reffig{bonham}.
Such a direct comparison is possible, since the experimental data has been brought to the absolute scale by normalizing the data using analytically known Bethe sum-rules~\cite{gosd:inok71}, with an reported average relative accuracy of around $2\%$.
Despite the application of the rather crude fixed-nuclei approximation, an overall good agreement between both theory and experiment can be seen for most values of $E$ and $\vartheta_p$, featuring a general tendency for the present results to underestimate both the position of the Bethe ridge's peak and the overall experimental data for large values of energy loss.
Supplemental to the CI results that use the relation \refeq{egosd} between GOSD and DDCS, an alternative description which includes an approximate correction factor for relativistic and exchange effects of the projectile may also be considered.
In order to provide an estimation for its influence on the results, the DDCS, as determined by using Eq. (2) in Ref.~\cite{gosd:ulsh74}, is shown in \reffig{bonham} in comparison to the uncorrected relation in \refeq{egosd}.
Its effect is to shift the theoretical results up by a few percent, but does not appear to contribute to an overall better agreement between the experimental data and the present results, as the theoretical over-estimations at low energies are even exacerbated in the same way.

The prominent resonant features visible only in the theoretical results below around 55 eV are artifacts caused by the neglect of nuclear motion in this work, which are known from investigations on photoionization to be broadened beyond visibility if vibronic degrees of freedom are properly taken into account.
Notably, there is a particularly large discrepancy between both results for higher energy values at $\vartheta_p = 4^\circ$, where the present results significantly underestimate the experimental ones.
This deviation is currently of unknown origin and can be hardly attributed to a systematic convergence deficiency, as the agreement of the DDCS for both smaller and larger values of $E$ and/or $\vartheta_p$ can generally be seen to be better.
Likewise, a potentially inaccurate normalization of the experimental data, which uses the Thomas-Reiche-Kuhn sum rule in order to bring the data to the absolute scale, cannot be attributed to these deviations, because, as for most angles, the present results typically already agree significantly better at lower energies.

The curves shown in \reffig{zurales} compare the above theoretical and experimental results to the available theoretical data of Zurales and Lucchese~\cite{gosd:zura88}.
While also adopting the fixed-nuclei approximation within a first-Born treatment, they differ from this work by choosing an approach based on a restricted-Hartree-Fock computation for the initial state and the distorted (DWLA, DWVA) and Coulomb-wave (CWLA, CWVA) approximations for length and velocity forms in the continuum, respectively.
Additionally, in contrast to the present results and those given in Ref.~\cite{gosd:ulsh74}, the data of Zurales and Lucchese are not provided in terms of fixed scattering angle $\vartheta_p$, but for fixed values of momentum transfer $K$ instead.
As neither of these quantities stays constant with respect to the other over a range of energy losses, their mutual assignment is only approximate and care must be taken when comparing at the percent level, particularly close to the slopes of the Bethe ridge where the DDCS are most-sensitive to variations of $K$.

With these caveats in mind, for the high-energy part depicted in \reffig{zurales_high}, the results appear to be in rather close agreement to each other, with no clear preference of the experimental outcome to lie closer to one of the theoretical results over the other.
The overall discrepancy, however, appears to grow for larger values of energy loss, for which the theoretical results give similar results.
As has been noted by Zurales and Lucchese, the effect of distortion on the Coulomb wave-like solutions for the ejected electron is almost invisible in this regime, but achieving good agreement to Ulsh \emph{et al.} was only possible in the velocity form.
The reason for the poorer agreement in length form has been attributed to its larger sensitivity towards electronic correlation in the initial state, while final-state correlation is less important for large energies, hence rendering the velocity form more accurate in this regime.
This is in contrast to the present results, which are entirely given in the length form and yet show comparable agreement to  the experiment of Ulsh \emph{et al.}.
Following the argument of Ref.~\cite{gosd:zura88}, the comparably large configuration series (\emph{cf.}\ \reftbl{gsci}) may already provide a sufficient amount of electronic initial-state correlation, in contrast to the Hartree-Fock description of the initial state employed in Ref.~\cite{gosd:zura88}, rendering differences between length and velocity forms marginal.
However, the outcomes differ more strongly for lower values of energy at closer proximity to the threshold, as can be seen in \reffig{zurales_low}.
The results of Zurales and Lucchese, despite being a fixed-nuclei description as well, do not show the resonance structures that are visible in the present CI data, which is just the result of ignoring the electronic correlations within their chosen final states.
While the CI outcome agrees rather well (and better than Ref.~\cite{gosd:zura88}) with the experimental data beyond 55 eV, it appears to overestimate the height of the emerging experimental Bethe ridge to an extent similar to that of the length-form results of Ref.~\cite{gosd:zura88} employing Coulomb or distorted waves.
Still, the better inclusion of electronic correlation in the CI result allows for a sufficiently large CI computation to reasonably match to the experimental data, whereas the agreement of the Coulomb and distorted-wave results additionally depend on the chosen form.
Coincidentally, for the given value of scattering angle, the resonances seen in the present results are located around the peak of the Bethe ridge between 30 and 40 eV, rendering it difficult to reliably assess the agreement for the peak position.
A second, single-channel computation which neglects any channel couplings lacks these resonance structures and is shown in \reffig{zurales_low} as well.
Both its peak position and height mostly resemble to the CWLA results, but converge towards the CI, DWVA, and CWVA results at higher energies, where the length-form results of Zurales and Lucchese are less accurate.
As the single-channel CI results used the same, strongly correlated ground-state description as is used for the fully coupled result, it can be seen that it is the initial and not the final-state correlations the length-form results are sensitive to.
A more involved computation involving nuclear motion is expected to dilute the resonance line-shapes, which would allow for a more reliable comparison to experiment at these particular scattering angles.

\begin{figure}
  \includegraphics[width=0.48\textwidth]{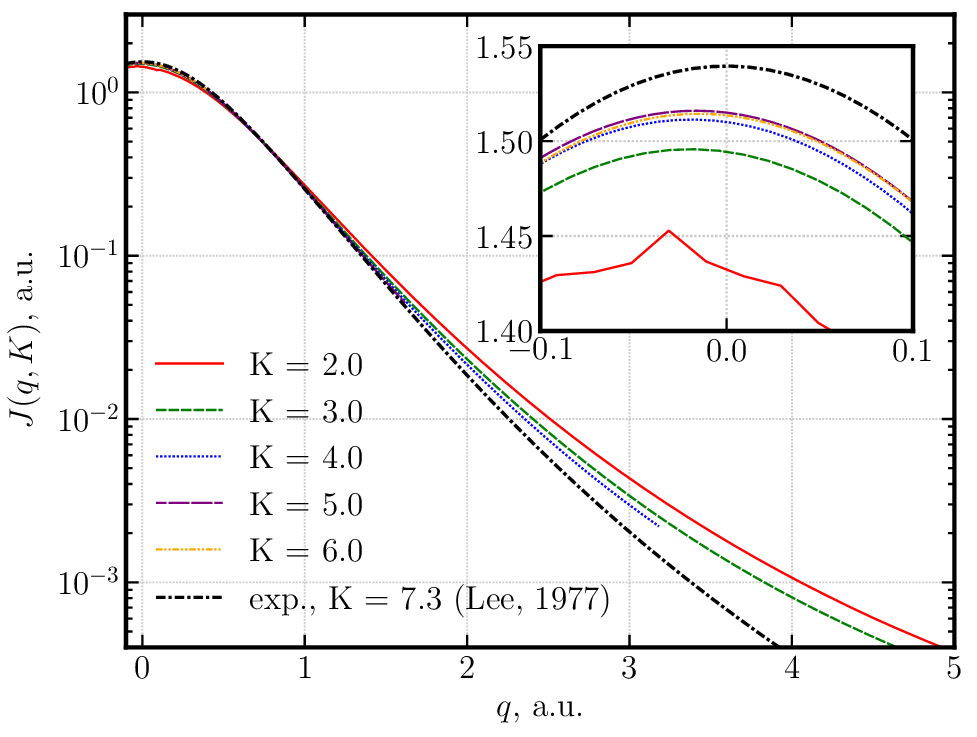}
  \caption{Compton profiles, according to \refeq{compprof}, on a logarithmic scale.
  The solid lines show the CI results for increasing values of momentum transfer in units of $a_0{}^{-1}$, as indicated in the legend.
  The black, dash-dotted curve depicts the fit for the experimental result of Lee, as given by Eq. (10) and Table VII in Ref.~\cite{gosd:lee77}.
  The inset provides a view on the region around the Compton peak in greater detail.}
  \labfig{compton}
\end{figure}

\subsection{Compton profiles}

Finally, it is possible to assess the quality of the Bethe-ridge description in the high-energy region by considering the effective Compton profiles as calculated from \refeq{compprof}.
In \reffig{compton}, the Compton profiles for several values of $K$ are presented and compared to the fitted experimental results of Lee~\cite{gosd:lee77}, who used a formula that fixes the Compton profile peak position to $q=0$ as is the case for the BEA/IA limit \refeq{beacomp}.
Although the highest reachable energy loss of around 560 eV is not quite large enough to permit a direct comparison to the experimental reference at $K=7.3$ $a_0{}^{-1}$, the overall agreement is nevertheless reasonable on the given scale.
Most notably, for larger values of $q$, the theoretical curves appear to approach the experimental outcome with increasing values of $K$.
Additionally, estimations using the BEA-form Compton profile \refeq{beacomp} have shown~\cite{gosd:ulsh72} that proper inclusion of vibronic motion could, at least in part, make up for the discrepancy seen with regard to the experimental Compton peak heights~\cite{gosd:eise70a, gosd:lee77}.
Thorough inclusion of vibrational averaging~\cite{gosd:ulsh72, gosd:smit77, gosd:smit80} raises the fixed-nuclei Compton-peak height at equilibrium $R=1.4$ $a_0$ by about 0.95\%, which partly explains the observed discrepancy.
In order to further explain the remaining discrepancy, it is necessary to dissect the numerical convergence errors from the breakdown of the BEA/IA at small values of $K$, \emph{i.\,e.}, the Compton defect.
Fortunately, Lee provided further data points for the Compton-peak heights at lower values of $K$~\cite{gosd:lee77}, where direct comparison is possible.
These are being presented, together with the theoretical results of Zurales and Lucchese~\cite{gosd:zura88}, in \reffig{comppeaks}.
The present vibration-adjusted CI peak-heights can be seen to agree very well to the experimental results, but tend to decrease after around $K=5$ $a_0{}^{-1}$, as observed before in \reffig{compton}.
The theoretical results of Zurales and Lucchese~\cite{gosd:zura88} show the length-form results to overestimate the experiment, whereas the DWVA is agreeing better with the CI and experimental results, particularly at lower values of $K$.

\begin{figure*}[t]
  \subfigure[][]{
  \includegraphics[width=0.48\textwidth]{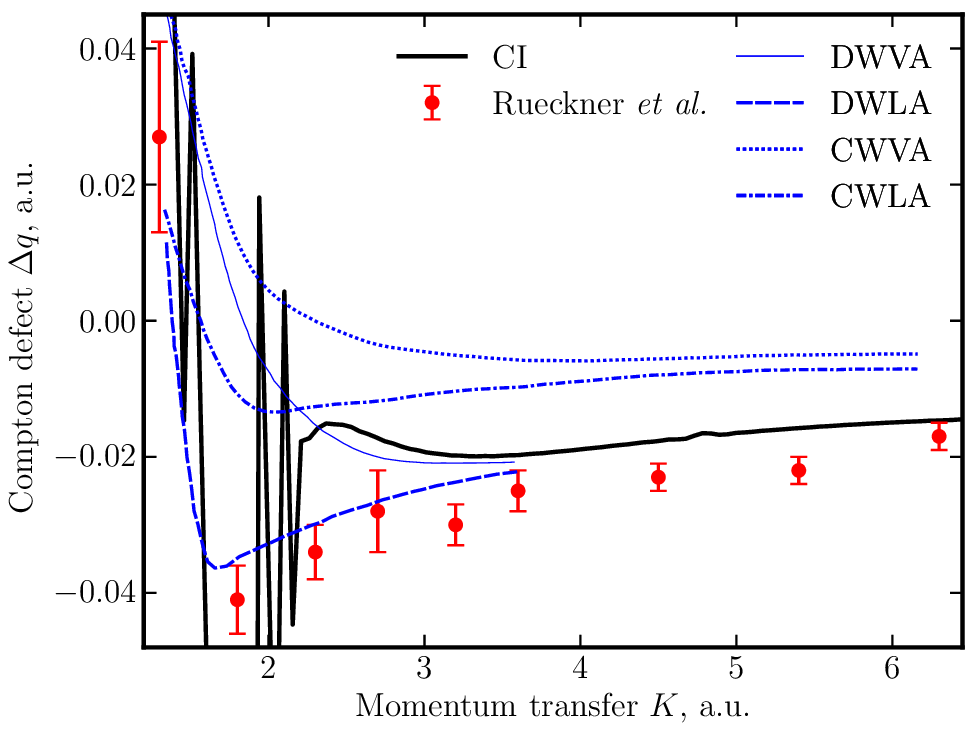}
  \labfig{compdefects}}
  \subfigure[][]{
  \includegraphics[width=0.48\textwidth]{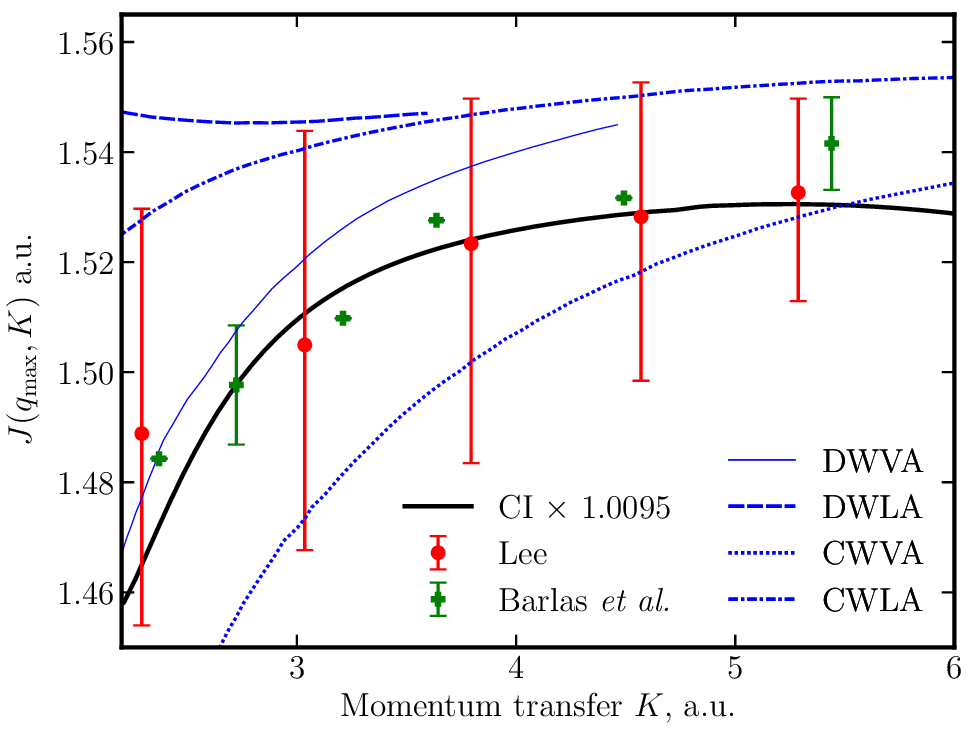}
  \labfig{comppeaks}}
  \caption{Comparison of the Compton defects $\Delta q=q_\mrm{max}$ \subref{fig:compdefects} and the peak heights \subref{fig:comppeaks} to experimental (red or green data points for Lee~\cite{gosd:lee77} and Rueckner \emph{et al.}~\cite{gosd:ruec78}, respectively) and theoretical results of Zurales and Lucchese~\cite{gosd:zura88} (blue thin, broken lines).
  The thick black line represents the present CI result.
  The profile heights in \subref{fig:comppeaks} are increased by 0.95\% in order to compensate for the disregard of nuclear motion (\emph{cf.}\ the discussion in the main text).}
  \labfig{compton_comp}
\end{figure*}

The Compton defect of the peak position $q_\mrm{max}$ is further compared to the experimental high-precision results of Rueckner \emph{et al.}~\cite{gosd:ruec78} in \reffig{compdefects}.
In order to estimate the peak position, the Compton profile has been fitted to parabolas using three points around the maximum value at the computed grid.
While the CI result is strongly superimposed by the autoionization line-shapes at lower values of $K$ (\emph{cf.}\ \reffig{zurales_low}), it tends to resemble the experimental outcome more closely at larger values of $K$, albeit missing the experimental error bars by a slight margin.
The inclusion of distortion to the scattering-state description in the DWVA and DWLA results of Ref.~\cite{gosd:zura88} improves the agreement to experiment, indicating the Compton-defect position to be more sensitive to the quality of the final states than the Compton-peak heights.
The authors of Ref.~\cite{gosd:zura88} corroborate this assessment by highlighting the good agreement of the DWLA result, which however is in contrast to what can be seen \reffig{comppeaks}, where the DWLA is performing worse.
Yet, the present CI results are expressed in length form as well and, surprisingly, show less agreement to experiment between $K=2.5$ and $3.0$ $a_0{}^{-1}$ (where the resonance effects are less pronounced), despite the arguably better initial and final-state description.
This might seem surprising, as it is known that the quality of the BEA as well as that of its corrections is not only sensitive to proper inclusion of final-state interactions between the ejected electron and the ion alone, but rather to the subtle cancellation of the interaction potential affecting both the initial and final states~\cite{gosd:holm89}.
As such, it may not be excluded that the good agreement of the DWLA result to the experiment for the Compton defect is, in fact, at least partially coincidental.

\section{Discussion}\labsec{disc} 

In strict terms, the description of the multichannel-scattering process as given in \refsec{fbm} is formally only justified for energies below the highest threshold of the ionic states included in the summation of \refeq{mc_asymp}.
This is also the case for full breakup (or double ionization) processes, which has not been resolved in this work.
Still, it is known from photon~\cite{gosd:fojo04, gosd:taka14} and electron-impact ionization experiments~\cite{gosd:koss90} and theory~\cite{gosd:weck07, gosd:zhu23} that the scattering amplitudes for both, higher excited channels and double ionization, are orders of magnitude smaller than the contributions of the few dominant channels.
The agreement of the CI results with the experimental data of Ulsh \emph{et al.}~\cite{gosd:ulsh74} also appears to corroborate this assessment.
Thus, the remaining discrepancies to the experimental results of Ulsh \emph{et al.}~\cite{gosd:ulsh74} and Lee~\cite{gosd:lee77} must originate from different causes.
Firstly, particularly for low ejection energies, a better description of nuclear motion is essential for the low-energy part of the Bethe surface.
As is the case for photoionization, the expected broadening of the Fano resonances is a minimal requirement to bring the theoretical results to a closer, \emph{qualitative} agreement.
Likewise, any relativistic and exchange effect has been largely neglected in the present work, although the impact of corrections introduced by the approximate factors as given by Refs.~\cite{gosd:bonh73} and~\cite{gosd:zura88} appear to be rather small and hence seem to be, at best, only partly responsible for the discrepancies observed in Figs. \ref{fig:bonham} and \ref{fig:zurales}.
However, it should also be kept in mind that the validity of such approximate correction factors is, to a certain degree, tied to that of the BEA, which is known to behave rather poorly for the larger values of $K$ considered in this work~\cite{gosd:lee77}.
The differences in the DDCS to the experiment as observed by Zurales and Lucchese~\cite{gosd:zura88} has been, in part, attributed to a lack of electronic correlation in the description of initial and/or final states.
Still, particularly close to threshold, deviations between theory and experiment prevail, albeit the present work features a more elaborate inclusion of electronic correlation, made possible by the virtue of better computational capabilities if nothing else.
The same might be concluded for the effects due to the non-spherical two-center character of the molecule, appearing to play a minor role only visible when discussing, \emph{e.\,g.}, the Compton defect.
The usage of prolate spheroidal coordinates provides the most natural description for such a two-center system, promising quicker convergence than, for instance, single-center approaches, and allowing for ruling out any single-center deficiencies from the list of uncertainties.
Taking above considerations into account, one thus may argue that better agreement might be achieved only in the context of inclusion of second Born amplitudes~\cite{gosd:houa03}, a distorted-wave description of the projectile, or some other formalism including post-collision interaction between the ejected electron and the projectile.

The tendency of the computed Compton-profile peaks to seemingly decrease for growing values of $K$ beyond $5.0$ $a_0{}^{-1}$, as observed in Figs. \ref{fig:compton} and \ref{fig:comppeaks}, is unexpected, as it is not in line with the findings of Zurales, Lucchese~\cite{gosd:zura88} and Lee~\cite{gosd:lee77}, which all show an, albeit slow, but convergent and monotonically increasing behavior in $K$.
Reducing both the cutoff number $\Lambda_\mrm{max}$ as well as $L_\mrm{max}$ showed to exacerbate this phenomenon at even smaller values of $K$.
Thus, these deviations are likely attributable to numerical convergence, in particular with respect to the number of quasi-angular channels (\emph{cf.}\ \reffig{lconv}), rather than to a deficiency in the model itself.
In particular, the systematic behavior of the partial-wave contributions for larger values of $L$ and $M$ as seen in \reffig{convergence} indicates that the strong centrifugal potential barrier overshadows properties of the molecular potential.
Therefore, it is conceivable that these higher partial-wave contributions might be estimated by single-center Coulomb waves rather than the numerically much more demanding scattering states \refeq{psich}.

A further source of numerical error lies in the thorough description of the initial-state correlation, which is not necessarily directly visible in the correlation energy alone.
For instance, photoionization cross-sections computed using the same initial and final CI states with formally equivalent length and velocity forms showed a relative discrepancy of around 2\% to 5\%~\cite{gosd:volk26a}.
Whether such a discrepancy also prevails for the GOS(D), for which an analogous velocity form can be given as well, remains an open question.
Either way, it is possible to significantly improve the initial-state representation of the present method by allowing the CI to be constructed from ionic orbitals living inside a much smaller numerical box than those used for representing the scattering continuum.
The loss of orthogonality between initial and final-state orbitals adds some complexity of computing additional overlap integrals, which, fortunately, is not the bottleneck of the present approach and poses no principal obstacle.
Likewise, a calculation of the BEA Compton-profile \refeq{beacomp} by employing the Fourier transform of the CI initial-state's one-particle density might represent an interesting internal-consistency cross-check, to which the present results are expected to converge in the large $K$ limit.
It is, however, still not entirely predictable for which values of $K$ the BEA/IA actually begins to be sufficiently accurate, as previous estimations have proven to be too optimistic in the past~\cite{gosd:bell86}.
As such, computations at higher values of $K$, particularly at the value of $7.2$ $a_0{}^{-1}$, might be interesting, as more experimental~\cite{gosd:lee77, gosd:ruec78} and theoretical~\cite{gosd:roet95} comparisons with regard to the Compton defect are possible.

Overall, the present prolate spheroidal free-boundary description has shown to be capable of providing fairly accurate doubly differential cross-section data.
While the full potential of this approach only truly unfolds at larger internuclear separations than the one considered in this work, it nonetheless gave a proof of principle for such a two-center description to be a viable pathway.
Scenarios in which scattering takes place in hotter gases, where there is a considerable amount of population residing in higher electronic states, may be a suitable further area of application for such a prolate spheroidal description.
This might be of particular relevance in areas of plasma and astrophysics.
Likewise, the FB method might also prove to be useful for problems involving multiple coupled continua, as, for instance, has been discussed for the case for alkali dimers~\cite{gosd:dumi07}.
Finally, as the FBA is applicable not only for electrons serving as projectiles, but also for other charged particles such as protons, positrons, or antiprotons, the presented Bethe surface might prove to be of use for revealing the significance of exchange effects at ejection energies where the BEA is breaking down.

\section{Summary}\labsec{summary}

In this work, the Bethe surface obtained from a theoretical computation of the generalized oscillator-strength density for randomly oriented H$_2$ molecules, up to a momentum transfer $K=8.0$ $a_0{}^{-1}$ and energy transfer of $E=558$ eV, has been presented.
Based on a prolate-spheroidal two-center CI description using ionic orbitals, the multi-channel free-boundary method allowed for the computation of scattering states, which include electronic-correlation effects such as autoionization.
In order to further assess the robustness of the method and to estimate the accuracy of the results, numerical convergence has been pushed systematically to the few-percent regime.
The obtained doubly-differential cross sections were found to be in good agreement to the experimental data of Ulsh \emph{et al.}~\cite{gosd:ulsh74} as well as to the theoretical findings of Zurales and Lucchese~\cite{gosd:zura88}.
Computations of the Compton profiles bear a close resemblance to the experimental result of Lee~\cite{gosd:lee77} and could partially reproduce the Compton defect, but existing discrepancies demonstrated the challenges connected to this method when faithful values of the Compton profile are sought for.
All in all, the obtained results contributed to shed light on the role of electronic correlation for both initial and final states in electron-impact ionization of molecular hydrogen as well as the validity of the BEA and are expected to be of direct future use for the energy-loss analysis of $\upbeta$-electron scattering on tritium molecules in the KATRIN~\cite{gosd:aker25} and TRISTAN~\cite{gosd:mert19} neutrino-mass experiments.

\subsection*{Acknowledgements}
The authors would like to express their gratitude towards Prof. Decleva for an insightful discussion.

\subsection*{Author contributions}
H.\ V.\ implemented the method, performed the calculations and drafted the manuscript. A.\ S.\ supervised the work. Both authors contributed equally to the evaluation and discussion of the results. 

\subsection{Data availability}
The data used to generate \reffig{bethe} are available from Ref.~\cite{Note2}.

%\vspace{1cm}

\appendix
\section{Molecular symmetry}\labsec{sym} 

A molecular single-electron symmetry $\gamma=(m,\wp)$ is defined by the transformation under $\hat z$-axis rotation and inversion w.\,r.\,t.\  the molecular center $\mbf r\to -\mbf r$.
The respective quantum numbers $\sigma,\pm\pi,\pm\delta,\ldots$ for $m=0, \pm 1, \pm2,\ldots$ (magnetic quantum number) and $\wp\in\{\mrm g, \mrm u\}$, with $\mrm g=1$ and $\mrm u=-1$ (parity) then may label any single-electron, that is, ionic state of index $n$, $\phi^\gamma_n(\mbf r\,;\, R)$.

Furthermore, let
\begin{equation}\labeq{sym}
  \Gamma \;=\; {}^{2S+1}\Lambda_{\mathcal P}^{(\pm)}
\end{equation}
denote the two-electron molecular symmetry.
It may be formed from two given single-particle symmetries $\gamma=(m,\wp)$ and $\bar\gamma=(\bar m,\bar\wp)$ with $\Lambda = |m+\bar m|=\Sigma,\Pi,\Delta,\ldots$ being the total angular momentum in $\hat z$ direction, $\mathcal P=\wp\bar\wp=\pm 1$ (\emph{gerade/ungerade}) the total parity, $S=0, 1$ (singlet/triplet multiplicity) the total spin, and $\pm$ the reflection symmetry upon a plane containing the internuclear axis (for $\Sigma$ states only).
While such a representation of the total symmetry is by no means the only one possible, it is the most natural description for an approach where the two-electron state is either given as a CI series of independent configurations or for asymptotically large separations, where correlation plays no role.

\bibliography{journals_v2,gosd}

@book{gosd:lapa,
      AUTHOR = {Anderson, E. and Bai, Z. and Bischof, C. and
                Blackford, S. and Demmel, J. and Dongarra, J. and
                Du Croz, J. and Greenbaum, A. and Hammarling, S. and
                McKenney, A. and Sorensen, D.},
      TITLE = {{LAPACK} Users' Guide},
      EDITION = {Third},
      PUBLISHER = {Society for Industrial and Applied Mathematics},
      YEAR = {1999},
      ADDRESS = {Philadelphia, PA},
      ISBN = {0-89871-447-8 (paperback)}
}

@book{gosd:flam57,
  address = {Palo Alto, Calif.},
  author = {Flammer, C.},
  publisher = {Stanford U. P.},
  title = {{S}pheroidal {W}ave {F}unctions},
  year = {1957}
}

@incollection{gosd:arri90,
  address = {New York},
  author = {Arrighini, G.~P. and Guidotti, C. and Durante, N.},
  booktitle = {Nonequilibrium Processes in Partially Ionized Gases},
  doi = {10.1007/978-1-4615-3780-9\_16},
  pages = {269--281},
  publisher = {Plenum Press},
  title = {{I}nelastic {S}cattering of {E}lectrons from $\mathrm{H}_2$ {M}olecule and {F}irst-{B}orn {A}pproximation: {R}ole of {C}orrelation},
  year = {1990}
}

@incollection{gosd:star23,
  address = {New York, NY},
  author = {Starace, A.~F.},
  booktitle = {Springer Handbook of Atomic, Molecu\-lar, and Optical Physics},
  doi = {10.1007/978-0-387-26308-3\_24},
  pages = {379--390},
  publisher = {Springer New York},
  title = {{P}hotoionization of {A}toms},
  year = {2006},
  editor = {Drake, G.},
}

@article{gosd:volk26a,
  author = {Volkmann, H. and Sch{\"u}rmann, J. and Saenz, A.},
  title  = {Accurate theoretical methods for photoionization of {H}$_2$ molecules},
  journal = {},
  archivePrefix = {arXiv},
  eprint = {2609.02461},
  primaryClass = {quant-ph},
  year   = {2026}
}

@article{gosd:beth30,
  author  = {Bethe, H.},
  doi     = {10.1002/andp.19303970303},
  journal = adp,
  pages   = {325--400},
  title   = {{Z}ur {T}heorie des {D}urchgangs schneller {K}orpuskularstrahlen durch {M}aterie},
  volume  = {397},
  year    = {1930}
}

@article{gosd:rued56,
  author  = {Ruedenberg, K. and Roothaan, C.~C.~J. and Jaunzemis, W.},
  doi     = {10.1063/1.1742457},
  journal = jcp,
  pages   = {201--220},
  title   = {{S}tudy of {T}wo-{C}enter {I}ntegrals {U}seful in {C}alculations on {M}olecular {S}tructure. {III}. {A} {U}nified {T}reatment of the {H}ybrid, {C}oulomb, and {O}ne-{E}lectron {I}ntegrals},
  volume  = {24},
  year    = {1956}
}

@article{gosd:boer61,
  author  = {Boersch, H. and Geiger, J. and Reich, H.~J.},
  doi     = {10.1007/BF01338936},
  journal = zp,
  pages   = {296--309},
  title   = {{E}nergieverluste von 25 ke{V}-{E}lektronen in atomarem {W}asserstoff},
  volume  = {161},
  year    = {1961}
}

@article{gosd:iiji63,
  author  = {Iijima, T. and Bonham, R.~A. and Ando, T.},
  doi     = {10.1021/j100801a017},
  journal = jcp,
  pages   = {1472--1474},
  title   = {{T}he {T}heory of {E}electron {S}cattering {F}rom {M}olecules {I}. {T}heoretical {D}evelopment},
  volume  = {67},
  year    = {1963}
}

@article{gosd:read63,
  author  = {Read, F.~H. and Whiterod, G.~L.},
  doi     = {10.1088/0370-1328/82/3/315},
  journal = pps,
  pages   = {434},
  title   = {{E}lectron {I}mpact {S}pectroscopy {I}. {T}he {D}etermination of the {S}ymmetry {S}pecies of {M}olecular {E}xcited {S}tates},
  volume  = {82},
  year    = {1963}
}

@article{gosd:geig64,
  author  = {Geiger, J.},
  doi     = {10.1007/bf01380873},
  journal = zp,
  pages   = {413--425},
  title   = {{S}treuung von 25 ke{V}-{E}lektronen an {G}asen: {III}. {S}treuung an molekularem {W}asserstoff},
  volume  = {181},
  year    = {1964}
}

@article{gosd:kolo65,
  author  = {Ko{\l}os, W. and Wolniewicz, L.},
  doi     = {10.1016/0022-2852(76)90281-2},
  journal = jcp,
  pages   = {2429--2441},
  title   = {{P}otential-{E}nergy {C}urves for the {{X}${}^1\Sigma_g^+$}, b{${}^3\Sigma_u^+$}, and {{C}${}^1\Pi_u$} {S}tates of the {H}ydrogen {M}olecule},
  volume  = {43},
  year    = {1965}
}

@article{gosd:cart67,
  author  = {Cartwright, D.~C. and Kuppermann, A.},
  doi     = {10.1103/physrev.163.86},
  journal = pr,
  pages   = {86--102},
  title   = {{E}lectron-{I}mpact {E}xcitation {C}ross {S}ection for the {T}wo {L}owest {T}riplet {S}tates of {M}olecular {H}ydrogen},
  volume  = {163},
  year    = {1967}
}

@article{gosd:vrie68,
  author  = {Vriens, L. and Bonsen, T.~F.~M.},
  doi     = {10.1088/0022-3700/1/6/316},
  journal = jpb,
  pages   = {1123},
  title   = {{D}ifferential cross sections for ionization of the hydrogen atom by fast charged particles in the binary-encounter theory and bethe theory},
  volume  = {1},
  year    = {1968}
}

@article{gosd:mehl69,
  author  = {Mehler, E.~L. and Ruedenberg, K.},
  doi     = {10.1063/1.1671417},
  journal = jcp,
  pages   = {2575--2580},
  title   = {{T}wo-{C}enter {E}xchange {I}ntegrals between {S}later-{T}ype {A}tomic {O}rbitals},
  volume  = {50},
  year    = {1969}
}

@article{gosd:eise70,
  author  = {Eisenberger, P. and Platzman, P.~M.},
  doi     = {10.1103/physreva.2.415},
  journal = pra,
  pages   = {415--423},
  title   = {{C}ompton {S}cattering of {X} {R}ays from {B}ound {E}lectrons},
  volume  = {2},
  year    = {1970}
}

@article{gosd:eise70a,
  author  = {Eisenberger, P.},
  doi     = {10.1103/physreva.2.1678},
  journal = pra,
  pages   = {1678--1686},
  title   = {{E}lectron {M}omentum {D}ensity of {H}e and {H}$_{2}$; {C}ompton {X}-{R}ay {S}cattering},
  volume  = {2},
  year    = {1970}
}

@article{gosd:agui71,
  author  = {Aguilar, J. and Combes, J.~M.},
  doi     = {10.1007/bf01877510},
  journal = cmp,
  pages   = {269--279},
  title   = {{A} class of analytic perturbations for one-body {S}chr{\"o}dinger {H}amiltonians},
  volume  = {22},
  year    = {1971}
}

@article{gosd:bals71,
  author  = {Balslev, E. and Combes, J.~M.},
  doi     = {10.1007/bf01877511},
  journal = cmp,
  pages   = {280--294},
  title   = {{S}pectral properties of many-body {S}chr{\"o}dinger operators with dilatation-analytic interactions},
  volume  = {22},
  year    = {1971}
}

@article{gosd:inok71,
  author  = {Inokuti, M.},
  doi     = {10.1103/revmodphys.43.297},
  journal = rmp,
  pages   = {297--347},
  title   = {{I}nelastic {C}ollisions of {F}ast {C}harged {P}articles with {A}toms and {M}olecules---{T}he {B}ethe {T}heory {R}evisited},
  volume  = {43},
  year    = {1971}
}

@article{gosd:simo72,
  author  = {Simon, B.},
  doi     = {10.1007/bf01649654},
  journal = cmp,
  pages   = {1--9},
  title   = {{Q}uadratic form techniques and the {B}alslev-{C}ombes theorem},
  volume  = {27},
  year    = {1972}
}

@article{gosd:ulsh72,
  author  = {Ulsh, R. and Bonham, R. and Bartell, L.},
  doi     = {10.1016/0009-2614(72)80029-0},
  journal = cpl,
  pages   = {6--8},
  title   = {{V}ibrational correction to the calculation of the {C}ompton profile of {H}$_2$},
  volume  = {13},
  year    = {1972}
}

@article{gosd:bonh73,
  author  = {Bonham, R.~A. and Tavard, C.},
  doi     = {10.1063/1.1680682},
  journal = jcp,
  pages   = {4691--4704},
  title   = {{Q}uantum mechanical first {B}orn binary encounter theory of electron impact ionization. {II} {E}lectron {C}ompton scattering},
  volume  = {59},
  year    = {1973}
}

@article{gosd:liu73,
  author  = {Liu, J.~W.},
  doi     = {10.1103/PhysRevA.7.103},
  journal = pra,
  pages   = {103--109},
  title   = {{T}otal {I}nelastic {C}ross {S}ection for {C}ollisions of {H}$_2$ with {F}ast {C}harged {P}articles},
  volume  = {7},
  year    = {1973}
}

@article{gosd:simo73,
  author  = {Simon, B.},
  doi     = {10.2307/1970847},
  journal = am,
  pages   = {247},
  title   = {{R}esonances in n-{B}ody {Q}uantum {S}ystems {W}ith {D}ilatation {A}nalytic {P}otentials and the {F}oundations of {T}ime-{D}ependent {P}erturbation {T}heory},
  volume  = {97},
  year    = {1973}
}

@article{gosd:ulsh74,
  author  = {Ulsh, R.~C. and Wellenstein, H.~F. and Bonham, R.~A.},
  doi     = {10.1063/1.1680755},
  journal = jcp,
  pages   = {103--111},
  title   = {{B}ethe surface, elastic and inelastic differential cross sections, {C}ompton profile, and binding effects for {H}$_2$ obtained by electron scattering with 25 ke{V} incident electrons},
  volume  = {60},
  year    = {1974}
}

@article{gosd:lee77,
  author  = {Lee, J.~S.},
  doi     = {10.1063/1.433829},
  journal = jcp,
  pages   = {4906--4914},
  title   = {{A}ccurate determination of {H}$_2$, {H}e, and {D}$_2$ {C}ompton profiles by high energy electron impact spectroscopy},
  volume  = {66},
  year    = {1977}
}

@article{gosd:smit77,
  author  = {Smith, V.~H. and Thakkar, A.~J. and Henneker, W.~H. and Liu, J.~W. and Liu, B. and Brown, R.~E.},
  doi     = {10.1063/1.435307},
  journal = jcp,
  pages   = {3676--3682},
  title   = {{A}ccurate {C}ompton profiles for {H}$_2$ and {D}$_2$ including the effects of electron correlation and molecular vibration and rotation},
  volume  = {67},
  year    = {1977}
}

@article{gosd:barl78,
  author  = {Barlas, A.~D. and Rueckner, W.~H.~E. and Wellenstein, H.~F.},
  doi     = {10.1088/0022-3700/11/19/014},
  journal = jpb,
  pages   = {3381},
  title   = {{A} critical evaluation of high energy electron impact spectroscopy to measure {C}ompton profiles},
  volume  = {11},
  year    = {1978}
}

@article{gosd:ruec78,
  author  = {Rueckner, W.~H.~E. and Barlas, A.~D. and Wellenstein, H.~F.},
  doi     = {10.1103/PhysRevA.18.895},
  journal = pra,
  pages   = {895--909},
  title   = {{E}lectron {C}ompton defect observed in {H}e, {H}$_2$, {D}$_2$, {N}$_2$, and {N}e profiles},
  volume  = {18},
  year    = {1978}
}

@article{gosd:arri80,
  author  = {Arrighini, G. and Biondi, F. and Guidotti, C.},
  doi     = {10.1080/00268978000103701},
  journal = mp,
  pages   = {1501--1514},
  title   = {{A} study of the inelastic scattering of fast electrons from molecular hydrogen},
  volume  = {41},
  year    = {1980}
}

@article{gosd:lane80,
  author  = {Lane, N.~F.},
  doi     = {10.1103/revmodphys.52.29},
  journal = rmp,
  pages   = {29--119},
  title   = {{T}he theory of electron-molecule collisions},
  volume  = {52},
  year    = {1980}
}

@article{gosd:smit80,
  author  = {Smith, V.~H. and Thakkar, A.~J. and Henneker, W.~H. and Liu, J.~W. and Liu, B. and Brown, R.~E.},
  doi     = {10.1063/1.440771},
  journal = jcp,
  pages   = {4150--4150},
  title   = {{E}rratum: {A}ccurate {C}ompton profiles for {H}$_2$ and {D}$_2$ including the effects of electron correlation and molecular vibration and rotation [{J}. {C}hem. {P}hys. \textbf{67}, 3676 (1977)]},
  volume  = {73},
  year    = {1980}
}

@article{gosd:kolo82,
  author  = {Ko{\l}os, W. and Monkhorst, H.~J. and Szalewicz, K.},
  doi     = {10.1063/1.443955},
  journal = jcp,
  pages   = {1323--1334},
  title   = {{E}nergy unresolved differential cross section for electron scattering by {H}$_2$},
  volume  = {77},
  year    = {1982}
}

@article{gosd:kolo83,
  author  = {Ko{\l}os, W{\l}odzimierz and Monkhorst, H.~J. and Szalewicz, K.},
  doi     = {10.1016/0092-640x(83)90016-5},
  journal = adndt,
  pages   = {239--263},
  title   = {{G}eneralized oscillator strengths for {X}-{B} transitions in the hydrogen molecule},
  volume  = {28},
  year    = {1983}
}

@article{gosd:leun83,
  author  = {Leung, K. and Brion, C.},
  doi     = {10.1016/0301-0104(83)85351-8},
  journal = cp,
  pages   = {113--137},
  title   = {{B}inary $(e, 2e)$ spectroscopic study and momentum space chemistry of the two-electron systems {H}e and {H}$_2$},
  volume  = {82},
  year    = {1983}
}

@article{gosd:froe85,
  author  = {Froelich, P. and Flores-Riveros, A. and Weyrich, W.},
  doi     = {10.1063/1.448326},
  journal = jcp,
  pages   = {2305--2312},
  title   = {{N}onrelativistic {C}ompton scattering in {F}urry’s picture. {II}. {B}ethe surface by means of the complex-coordinate method},
  volume  = {82},
  year    = {1985}
}

@article{gosd:bell86,
  author  = {Bell, F.},
  doi     = {10.1063/1.451656},
  journal = jcp,
  pages   = {303--307},
  title   = {{O}n the {C}ompton defect},
  volume  = {85},
  year    = {1986}
}

@article{gosd:bonh86,
  author  = {Bonham, R.~A. and Goruganthu, R.~R.},
  doi     = {10.1063/1.450288},
  journal = jcp,
  pages   = {3068--3077},
  title   = {{T}he second {B}orn approximation for electron scattering. {II}. {T}he high energy limit for small angle dipole allowed inelastic scattering from atoms},
  volume  = {84},
  year    = {1986}
}

@article{gosd:liu87,
  author  = {Liu, J.~W.},
  doi     = {10.1103/PhysRevA.35.591},
  journal = pra,
  pages   = {591--597},
  title   = {{T}otal cross sections for high-energy electron scattering by {H}$_2({}^1{\Sigma}^+_\mathrm{g})$, {N}$_2({}^1{\Sigma}^+_\mathrm{g})$, and {O}$_2({}^3{\Sigma}^+_g$)},
  volume  = {35},
  year    = {1987}
}

@article{gosd:lahm88,
  author  = {Lahmam-Bennani, A. and Avaldi, L. and Fainelli, E. and Stefani, G.},
  doi     = {10.1088/0953-4075/21/11/026},
  journal = jpb,
  pages   = {2145},
  title   = {{T}he asymmetric $(e, 2e)$ collisions at intermediate and high impact energy: success and limits of first-order models},
  volume  = {21},
  year    = {1988}
}

@article{gosd:zura88,
  author  = {Zurales, R.~W. and Lucchese, R.~R.},
  doi     = {10.1103/physreva.37.1176},
  journal = pra,
  pages   = {1176--1184},
  title   = {{D}ifferential cross sections for the electron-impact ionization of molecular hydrogen in the distorted-wave {B}orn approximation},
  volume  = {37},
  year    = {1988}
}

@article{gosd:cher89,
  author  = {Cherid, M. and Lahmam-Bennani, A. and Duguet, A. and Zurales, R.~W. and Lucchese, R.~R. and Cappello, M.~C.~D. and Cappello, C.~D.},
  doi     = {10.1088/0953-4075/22/21/012},
  journal = jpb,
  pages   = {3483},
  title   = {{T}riple differential cross sections for molecular hydrogen, both under {B}ethe ridge conditions and in the dipolar regime. {E}xperiments and theory},
  volume  = {22},
  year    = {1989}
}

@article{gosd:holm89,
  author  = {Holm, P. and Ribberfors, R.},
  doi     = {10.1103/PhysRevA.40.6251},
  journal = pra,
  pages   = {6251--6259},
  title   = {{F}irst correction to the nonrelativistic {C}ompton cross section in the impulse approximation},
  volume  = {40},
  year    = {1989}
}

@article{gosd:koss90,
  author  = {Kossmann, H. and Schwarzkopf, O. and Schmidt, V.},
  doi     = {10.1088/0953-4075/23/2/012},
  journal = jpb,
  pages   = {301},
  title   = {{A}bsolute ionisation cross sections for electron impact on {H}$_2$},
  volume  = {23},
  year    = {1990}
}

@article{gosd:mart91,
  author  = {Mart{\'i}n, F. and Riera, A. and S{\'a}nchez, I.},
  doi     = {10.1063/1.460613},
  journal = jcp,
  pages   = {4275--4281},
  title   = {{F}ully ${L}^2$ methods for multichannel scattering problems. {P}artial widths},
  volume  = {94},
  year    = {1991}
}

@article{gosd:mcca91,
  author  = {McCarthy, I.~E. and Weigold, E.},
  doi     = {10.1088/0034-4885/54/6/001},
  journal = rpp,
  pages   = {789},
  title   = {{E}lectron momentum spectroscopy of atoms and molecules},
  volume  = {54},
  year    = {1991}
}

@article{gosd:rudd91,
  author  = {Rudd, M.~E.},
  doi     = {10.1103/PhysRevA.44.1644},
  journal = pra,
  pages   = {1644--1652},
  title   = {{D}ifferential and total cross sections for ionization of helium and hydrogen by electrons},
  volume  = {44},
  year    = {1991}
}

@article{gosd:bros92,
  author  = {Brosolo, M. and Decleva, P.},
  doi     = {10.1016/0301-0104(92)80069-8},
  journal = cp,
  pages   = {185--196},
  title   = {{V}ariational approach to continuum orbitals in a spline basis: {A}n application to {H}$_2{}^+$ photoionization},
  volume  = {159},
  year    = {1992}
}

@article{gosd:bros92a,
  author  = {Brosolo, M. and Decleva, P. and Lisini, A.},
  doi     = {10.1016/0010-4655(92)90009-n},
  journal = cpc,
  pages   = {207--214},
  title   = {{C}ontinuum wavefunctions calculations with least-squares schemes in a {B}-splines basis},
  volume  = {71},
  year    = {1992}
}

@article{gosd:liu93,
  author  = {Liu, J.~W. and Hagstrom, S.},
  doi     = {10.1103/physreva.48.166},
  journal = pra,
  pages   = {166--172},
  title   = {{G}eneralized oscillator-strength calculations for some low-lying excited states of {H}$_{2}$ using a high-accuracy configuration-interaction wave function},
  volume  = {48},
  year    = {1993}
}

@article{gosd:saen93,
  author  = {Saenz, A. and Weyrich, W.},
  doi     = {10.1515/zna-1993-1-245},
  journal = zna,
  pages   = {243--250},
  title   = {{T}he {G}eneralised {O}scillator-{S}trength {D}ensity of the {H}elium {A}tom, {C}alculated by a {N}ew {I}mplementation of the {C}omplex-{C}oordinate {M}ethod},
  volume  = {48},
  year    = {1993}
}

@article{gosd:cort94,
  author  = {Cortes, M. and Martin, F.},
  doi     = {10.1088/0953-4075/27/23/017},
  journal = jpb,
  pages   = {5741},
  title   = {{M}ultichannel close-coupling method with ${L}^2$ integrable bases},
  volume  = {27},
  year    = {1994}
}

@article{gosd:kim94,
  author  = {Kim, Yong-Ki and Rudd, E.~M.},
  doi     = {10.1103/physreva.50.3954},
  journal = pra,
  pages   = {3954--3967},
  title   = {{B}inary-encounter-dipole model for electron-impact ionization},
  volume  = {50},
  year    = {1994}
}

@article{gosd:roet95,
  author  = {Roeth, M. and Gasser, F. and Tavard, C.},
  doi     = {10.1002/qua.560530513},
  journal = ijqc,
  pages   = {569--574},
  title   = {{A}symmetries and anisotropies in the {C}ompton scattering from the hydrogen molecule},
  volume  = {53},
  year    = {1995}
}

@article{gosd:saen96,
  author  = {Saenz, A. and Weyrich, W. and Froelich, P.},
  doi     = {10.1088/0953-4075/29/1/014},
  journal = jpb,
  pages   = {97},
  title   = {{T}he first {B}orn approximation and absolute scattering cross sections},
  volume  = {29},
  year    = {1996}
}

@article{gosd:lerm97,
  author  = {Lermer, N. and Todd, B.~R. and Cann, N.~M. and Zheng, Y. and Brion, C.~E. and Yang, Z. and Davidson, E.~R.},
  doi     = {10.1103/PhysRevA.56.1393},
  journal = pra,
  pages   = {1393--1402},
  title   = {{E}lectron momentum spectroscopy of {H}$_2$ and {D}$_2$: {I}onization to ground and excited final states},
  volume  = {56},
  year    = {1997}
}

@article{gosd:sanc97,
  author  = {S{\'a}nchez, I. and Mart{\'i}n, F.},
  doi     = {10.1088/0953-4075/30/3/021},
  journal = jpb,
  pages   = {679},
  title   = {{R}epresentation of the electronic continuum of $\mathrm{H}_2$ with {$B$}-spline basis},
  volume  = {30},
  year    = {1997}
}

@article{gosd:lamb98,
  author  = {Lambropoulos, P. and Maragakis, P. and Zhang, J.},
  doi     = {10.1016/s0370-1573(98)00027-1},
  journal = prp,
  pages   = {203--293},
  title   = {{T}wo-electron atoms in strong fields},
  volume  = {305},
  year    = {1998}
}

@article{gosd:borg99,
  author  = {Borges, I. and Bielschowsky, C.~E.},
  doi     = {10.1103/physreva.60.1226},
  journal = pra,
  pages   = {1226--1234},
  title   = {{P}hoton and high-energy--electron-impact vibronic excitation of molecular hydrogen},
  volume  = {60},
  year    = {1999}
}

@article{gosd:mart99,
  author  = {Mart{\'i}n, F.},
  doi     = {10.1088/0953-4075/32/16/201},
  journal = jpb,
  pages   = {R197},
  title   = {{I}onization and dissociation using {B}-splines: photoionization of the hydrogen molecule},
  volume  = {32},
  year    = {1999}
}

@article{gosd:neud99,
  author  = {Neudachin, V.~G. and Popov, Y.~V. and Smirnov, Y.~F.},
  doi     = {10.1070/PU1999v042n10ABEH000492},
  journal = {Phys.-Usp.},
  pages   = {1017--1044},
  title   = {{E}lectron momentum spectroscopy of atoms, molecules, and thin films},
  volume  = {42},
  year    = {1999}
}

@article{gosd:weck00,
  author  = {Weck, P. and Joulakian, B. and Hanssen, J. and Foj\'on, O.~A. and Rivarola, R.~D.},
  doi     = {10.1103/physreva.63.042709},
  journal = pra,
  pages   = {014701},
  title   = {{M}ultiple differential cross sections for single ionization of {H}$_{2},$ {D}$_{2},$ and {T}$_{2}$ molecules by fast electron impact: {I}nfluence of vibrational states},
  volume  = {62},
  year    = {2000}
}

@article{gosd:bach01,
  author  = {Bachau, H. and Cormier, E. and Decleva, P. and Hansen, J.~E. and Mart{\'i}n, F.},
  doi     = {10.1088/0034-4885/64/12/205},
  journal = rpp,
  pages   = {1815},
  title   = {{A}pplications of {B}-splines in atomic and molecular physics},
  volume  = {64},
  year    = {2001}
}

@article{gosd:star01,
  author  = {Staroverov, V.~N. and Davidson, E.~R.},
  doi     = {10.1080/00268970010007299},
  journal = mp,
  pages   = {175--186},
  title   = {\emph{{A}b initio} {C}ompton maps of small molecules},
  volume  = {99},
  year    = {2001}
}

@article{gosd:weck01,
  author  = {Weck, P. and Foj\'on, O.~A. and Hanssen, J. and Joulakian, B. and Rivarola, R.~D.},
  doi     = {10.1103/physreva.63.042709},
  journal = pra,
  pages   = {042709},
  title   = {{T}wo-effective center approximation for the single ionization of molecular hydrogen by fast electron impact},
  volume  = {63},
  year    = {2001}
}

@article{gosd:bure02,
  author  = {Van Buren, A.~L. and Boisvert, J.~E.},
  doi     = {10.1090/qam/1914443},
  journal = qam,
  pages   = {589--599},
  title   = {{A}ccurate calculation of prolate spheroidal radial functions of the first kind and their first derivatives},
  volume  = {60},
  year    = {2002}
}

@article{gosd:sero02,
  author  = {Serov, V.~V. and Joulakian, B.~B. and Pavlov, D.~V. and Puzynin, I.~V. and Vinitsky, S.~I.},
  doi     = {10.1103/physreva.65.062708},
  journal = pra,
  pages   = {062708},
  title   = {$(e,2e)$ ionization of {H}$_{2}{}^{+}$ by fast electron impact: {A}pplication of the exact nonrelativistic two-center continuum wave},
  volume  = {65},
  year    = {2002}
}

@article{gosd:houa03,
  author  = {Houamer, S. and Mansouri, A. and Cappello, C.~D. and Lahmam-Bennani, A. and Elazzouzi, S. and Moulay, M. and Charpentier, I.},
  doi     = {10.1088/0953-4075/36/14/304},
  journal = jpb,
  pages   = {3009},
  title   = {{S}econd {B}orn approximation for the ionization of {H}$_2$ by electron impact},
  volume  = {36},
  year    = {2003}
}

@article{gosd:kapl03,
  author  = {Kaplan, I.~G. and Barbiellini, B. and Bansil, A.},
  doi     = {10.1103/PhysRevB.68.235104},
  journal = prb,
  pages   = {235104},
  title   = {{C}ompton scattering beyond the impulse approximation},
  volume  = {68},
  year    = {2003}
}

@article{gosd:stia03,
  author  = {Stia, C.~R. and Foj{\'o}n, O.~A. and Weck, P.~F. and Hanssen, J. and Rivarola, R.~D.},
  doi     = {10.1088/0953-4075/36/17/101},
  journal = jpb,
  pages   = {L257},
  title   = {{I}nterference effects in single ionization of molecular hydrogen by electron impact},
  volume  = {36},
  year    = {2003}
}

@article{gosd:taka03,
  author  = {Takahashi, M. and Khajuria, Y. and Udagawa, Y.},
  doi     = {10.1103/PhysRevA.68.042710},
  journal = pra,
  pages   = {042710},
  title   = {$(e,2e)$ ionization-excitation of {H}$_2$},
  volume  = {68},
  year    = {2003}
}

@article{gosd:bure04,
  author  = {Van Buren, A.~L. and Boisvert, J.~E.},
  doi     = {10.1090/qam/2086042},
  journal = qam,
  pages   = {493--507},
  title   = {{I}mproved calculation of prolate spheroidal radial functions of the second kind and their first derivatives},
  volume  = {62},
  year    = {2004}
}

@article{gosd:fojo04,
  author  = {Foj{\'o}n, O.~A. and Fern{\'a}ndez, J. and Palacios, A. and Rivarola, R.~D. and Mart{\'i}n, F.},
  doi     = {10.1088/0953-4075/37/15/003},
  journal = jpb,
  pages   = {3035},
  title   = {{I}nterference effects in {H}$_2$ photoionization at high energies},
  volume  = {37},
  year    = {2004}
}

@article{gosd:vann04,
  author  = {Vanne, Y.~V. and Saenz, A.},
  doi     = {10.1088/0953-4075/37/20/005},
  journal = jpb,
  pages   = {4101},
  title   = {{N}umerical treatment of diatomic two-electron molecules using a {B}-spline based {CI} method},
  volume  = {37},
  year    = {2004}
}

@article{gosd:sero05,
  author  = {Serov, V.~V. and Joulakian, B.~B. and Derbov, V.~L. and Vinitsky, S.~I.},
  doi     = {10.1088/0953-4075/38/15/014},
  journal = jpb,
  pages   = {2765},
  title   = {{I}onization excitation of diatomic systems having two active electrons by fast electron impact: a probe to electron correlation},
  volume  = {38},
  year    = {2005}
}

@article{gosd:capp06,
  author  = {Dal Cappello, C. and Mansouri, A. and Houamer, S. and Joulakian, B.},
  doi     = {10.1088/0953-4075/39/11/009},
  journal = jpb,
  pages   = {2431},
  title   = {{S}econd-order effects in $(e, 2e)$ ionization-excitation of {H}$_2$},
  volume  = {39},
  year    = {2006}
}

@article{gosd:dumi07,
  author  = {Dumitriu, I. and Vanne, Y.~V. and Awasthi, M. and Saenz, A.},
  doi     = {10.1088/0953-4075/40/10/016},
  journal = jpb,
  pages   = {1821},
  title   = {{P}hotoionization of the alkali dimer cations {L}i$^+{}_2$, {N}a$^+{}_2$ and {L}i{N}a$^+$},
  volume  = {40},
  year    = {2007}
}

@article{gosd:weck07,
  author  = {Weck, P.~F.},
  doi     = {10.4208/cicp.2007.v2.p466},
  journal = {Commun.~Comput.~Phys.},
  pages   = {466--476},
  title   = {{D}ouble {I}onization of {M}olecular {H}ydrogen by {F}ast {E}lectron {I}mpact},
  volume  = {2},
  year    = {2007}
}

@article{gosd:sero09,
  author  = {Serov, V.~V. and Joulakian, B.~B.},
  doi     = {10.1103/physreva.80.062713},
  journal = pra,
  pages   = {062713},
  title   = {{I}mplementation of the external complex scaling method in spheroidal coordinates: {I}mpact ionization of molecular hydrogen},
  volume  = {80},
  year    = {2009}
}

@article{gosd:taka09,
  author  = {Takahashi, M.},
  doi     = {10.1246/bcsj.82.751},
  journal = bcsj,
  pages   = {751--777},
  title   = {{L}ooking at {M}olecular {O}rbitals in {T}hree-{D}imensional {F}orm: {F}rom {D}ream to {R}eality},
  volume  = {82},
  year    = {2009}
}

@article{gosd:prat10,
  author  = {Pratt, R. and LaJohn, L. and Florescu, V. and Suri{\' c}, T. and Chatterjee, B. and Roy, S.},
  doi     = {10.1016/j.radphyschem.2009.04.035},
  journal = rpc,
  pages   = {124--131},
  title   = {{C}ompton scattering revisited},
  volume  = {79},
  year    = {2010}
}

@article{gosd:yoon10,
  author  = {Yoon, Jung-Sik and Kim, Young-Woo and Kwon, Deuk-Chul and Song, Mi-Young and Chang, Won-Seok and Kim, Chang-Geun and Kumar, V. and Lee, B.},
  doi     = {10.1088/0034-4885/73/11/116401},
  journal = rpp,
  pages   = {116401},
  title   = {{E}lectron-impact cross sections for deuterated hydrogen and deuterium molecules},
  volume  = {73},
  year    = {2010}
}

@article{gosd:miya12,
  author  = {Miyagi, H. and Morishita, T. and Watanabe, S.},
  doi     = {10.1103/physreva.85.022708},
  journal = pra,
  pages   = {022708},
  title   = {{E}lectron scattering and photoionization of one-electron diatomic molecules},
  volume  = {85},
  year    = {2012}
}

@article{gosd:liu14,
  author  = {Liu, Ya-Wei and Mei, Xiao-Xun and Kang, X. and Yang, K. and Xu, Wei-Qing and Peng, Yi-Geng and Hiraoka, N. and Tsuei, Ku-Ding and Zhang, Peng-Fei and Zhu, Lin-Fan},
  doi     = {10.1103/PhysRevA.89.014502},
  journal = pra,
  pages   = {014502},
  title   = {{D}etermination of the electronic structure of atoms and molecules in the ground state: {M}easurement of molecular hydrogen by high-resolution x-ray scattering},
  volume  = {89},
  year    = {2014}
}

@article{gosd:taka14,
  author  = {Takahashi, K. and Sakata, Y. and Hino, Y. and Sakai, Y.},
  doi     = {10.1140/epjd/e2014-40703-9},
  journal = epjd,
  pages   = {83},
  title   = {{D}oubly excited states of molecular hydrogen by scattered electron-ion coincidence measurements},
  volume  = {68},
  year    = {2014}
}

@article{gosd:yuro14,
  author  = {Yurova, I.~Y. and Shevyakina, N.~K.},
  doi     = {10.1134/s1990793114010163},
  journal = {Russ.~J.~Phys.~Chem.~B},
  pages   = {1--8},
  title   = {{I}onization of diatomic molecules by electron impact},
  volume  = {8},
  year    = {2014}
}

@article{gosd:zhan14,
  author  = {Zhang, Z. and Shan, X. and Wang, T. and Wang, E. and Chen, X.},
  doi     = {10.1103/PhysRevLett.112.023204},
  journal = prl,
  pages   = {023204},
  title   = {{O}bservation of the {I}nterference {E}ffect in {V}ibrationally {R}esolved {E}lectron {M}omentum {S}pectroscopy of {H}$_2$},
  volume  = {112},
  year    = {2014}
}

@article{gosd:zhao15,
  author  = {Zhao, Xiao-Li and Yang, K. and Xu, Long-Quan and Ma, Yong-Peng and Yan, S. and Ni, Dong-Dong and Kang, X. and Liu, Ya-Wei and Zhu, Lin-Fan},
  doi     = {10.1088/1674-1056/24/3/033301},
  journal = chinpb,
  pages   = {033301},
  title   = {{C}ompton profile of molecular hydrogen},
  volume  = {24},
  year    = {2015}
}

@article{gosd:chad16,
  author  = {Chadney, J.~M. and Galand, M. and Koskinen, T.~T. and Miller, S. and Sanz-Forcada, J. and Unruh, Y.~C. and Yelle, R.~V.},
  doi     = {10.1051/0004-6361/201527442},
  journal = {A\&A},
  pages   = {A87},
  title   = {{EUV}-driven ionospheres and electron transport on extrasolar giant planets orbiting active stars},
  volume  = {587},
  year    = {2016}
}

@article{gosd:toff16,
  author  = {Toffoli, D. and Decleva, P.},
  doi     = {10.1021/acs.jctc.6b00627},
  journal = jctc,
  pages   = {4996--5008},
  title   = {{A} {M}ultichannel {L}east-{S}quares {B}-{S}pline {A}pproach to {M}olecular {P}hotoionization: {T}heory, {I}mplementation, and {A}pplications within the {C}onfiguration\textendash {I}nteraction {S}ingles {A}pproximation},
  volume  = {12},
  year    = {2016}
}

@article{gosd:mara17,
  author  = {Marante, C. and Klinker, M. and Corral, In{\'e}s and Gonz{\'a}lez-V{\'a}zquez, Jes{\'u}s and Argenti, L. and Mart{\'i}n, F.},
  doi     = {10.1021/acs.jctc.6b00907},
  journal = jctc,
  pages   = {499--514},
  title   = {{H}ybrid-{B}asis {C}lose-{C}oupling {I}nterface to {Q}uantum {C}hemistry {P}ackages for the {T}reatment of {I}onization {P}roblems},
  volume  = {13},
  year    = {2017}
}

@article{gosd:zamm17,
  author  = {Zammit, M.~C. and Fursa, D.~V. and Savage, J.~S. and Bray, I.},
  doi     = {10.1088/1361-6455/aa6e74},
  journal = jpb,
  pages   = {123001},
  title   = {{E}lectron\textendash and positron\textendash molecule scattering: development of the molecular convergent close-coupling method},
  volume  = {50},
  year    = {2017}
}

@article{gosd:li18,
  author  = {Li, X. and Ren, X. and Hossen, K. and Wang, E. and Chen, X. and Dorn, A.},
  doi     = {10.1103/PhysRevA.97.022706},
  journal = pra,
  pages   = {022706},
  title   = {{T}wo-center interference in electron-impact ionization of molecular hydrogen},
  volume  = {97},
  year    = {2018}
}

@article{gosd:tapl18,
  author  = {Tapley, J.~K. and Scarlett, L.~H. and Savage, J.~S. and Zammit, M.~C. and Fursa, D.~V. and Bray, I.},
  doi     = {10.1088/1361-6455/aac8fa},
  journal = jpb,
  pages   = {144007},
  title   = {{V}ibrationally resolved electron-impact excitation cross sections for singlet states of molecular hydrogen},
  volume  = {51},
  year    = {2018}
}

@article{gosd:xu18,
  author  = {Xu, Long-Quan and Kang, X. and Peng, Yi-Geng and Xu, X. and Liu, Ya-Wei and Wu, Y. and Yang, K. and Hiraoka, N. and Tsuei, Ku-Ding and Wang, Jian-Guo and Zhu, Lin-Fan},
  doi     = {10.1103/PhysRevA.97.032503},
  journal = pra,
  pages   = {032503},
  title   = {{C}omparative study of inelastic squared form factors of the vibronic states of ${B}^1{\Sigma}_\mathrm{u}^+$, ${C}^1{\Pi}_\mathrm{u}$, and ${EF}^1{\Sigma}_\mathrm{g}^+$ for molecular hydrogen: {I}nelastic x-ray and electron scattering},
  volume  = {97},
  year    = {2018}
}

@article{gosd:carr19,
  author  = {Moreno Carrascosa, Andr{\'e}s and Yong, H. and Crittenden, D.~L. and Weber, P.~M. and Kirrander, A.},
  doi     = {10.1021/acs.jctc.9b00056},
  journal = jctc,
  pages   = {2836--2846},
  title   = {{A}b {I}nitio {C}alculation of {T}otal {X}-ray {S}cattering from {M}olecules},
  volume  = {15},
  year    = {2019}
}

@article{gosd:klee19,
  author  = {Kleesiek, M. and Behrens, J. and Drexlin, G. and Eitel, K. and Erhard, M. and Formaggio, J.~A. and Gl{\"u}ck, F. and Groh, S. and H{\"o}tzel, M. and Mertens, S. and Poon, A.~W.~P. and Weinheimer, C. and Valerius, K.},
  doi     = {10.1140/epjc/s10052-019-6686-7},
  journal = epjc,
  pages   = {204},
  title   = {$\upbeta$-{D}ecay spectrum, response function and statistical model for neutrino mass measurements with the {KATRIN} experiment},
  volume  = {79},
  year    = {2019}
}

@article{gosd:mert19,
  author  = {Mertens, S. and Alborini, A. and Altenm{\"u}ller, K. and Bode, T. and Bombelli, L. and Brunst, T. and Carminati, M. and Fink, D. and Fiorini, C. and Houdy, T. and Huber, A. and Korzeczek, M. and Lasserre, T. and Lechner, P. and Manotti, M. and Peric, I. and Radford, D.~C. and Siegmann, D. and Slez{\'a}k, M. and Valerius, K. and Wolf, J. and W{\"u}stling, S.},
  doi     = {10.1088/1361-6471/ab12fe},
  journal = jpg,
  pages   = {065203},
  title   = {{A} novel detector system for {KATRIN} to search for ke{V}-scale sterile neutrinos},
  volume  = {46},
  year    = {2019}
}

@article{gosd:dhan20,
  author  = {Dhankhar, N. and Choubisa, R.},
  doi     = {10.1088/1361-6455/abcb52},
  journal = jpb,
  pages   = {015203},
  title   = {{E}lectron impact single ionization of hydrogen molecule by twisted electron beam},
  volume  = {54},
  year    = {2020}
}

@article{gosd:wang20,
  author  = {Wang, Shu-Xing and Zhu, Lin-Fan},
  doi     = {10.1063/5.0011416},
  journal = {Matter~Rad.~Extremes},
  pages   = {054201},
  title   = {{N}on-resonant inelastic {X}-ray scattering spectroscopy: {A} momentum probe to detect the electronic structures of atoms and molecules},
  volume  = {5},
  year    = {2020}
}

@article{gosd:aker21,
  author  = {Aker, M. and others},
  collaboration = {The KATRIN Collaboration},
  doi     = {10.1140/epjc/s10052-021-09325-z},
  journal = epjc,
  pages   = {579},
  title   = {{P}recision measurement of the electron energy-loss function in tritium and deuterium gas for the {KATRIN} experiment},
  volume  = {81},
  year    = {2021}
}

@article{gosd:onit22,
  author  = {Onitsuka, Y. and Tachibana, Y. and Takahashi, M.},
  doi     = {10.1039/d2cp02461f},
  journal = pccp,
  pages   = {19716--19721},
  title   = {{A}symptotic behavior of the electron-atom {C}ompton profile due to the intramolecular {H}-atom motion in {H}$_2$},
  volume  = {24},
  year    = {2022}
}

@article{gosd:sing23,
  author  = {Singor, A. and Jeremy Savage, S. and Bray, I. and Barry Schneider, I. and Dmitry Fursa, V.},
  doi     = {10.1016/j.cpc.2022.108514},
  journal = cpc,
  pages   = {108514},
  title   = {{C}ontinuum solutions to the two\textendash center {C}oulomb problem in prolate spheroidal coordinates},
  volume  = {282},
  year    = {2023}
}

@article{gosd:wan23,
  author  = {Wan, Jian-Jie and Gu, J. and Wu, Zhao-Yang and Wu, F. and Li, J. and Qiao, Hao-Xue},
  doi     = {10.1016/j.rinp.2023.106562},
  journal = {Results in Physics},
  pages   = {106562},
  title   = {{H}igh-accuracy calculation of nonrelativistic {C}ompton profile for {H}-like ions},
  volume  = {50},
  year    = {2023}
}

@article{gosd:zhu23,
  author  = {Zhu, Yu-Hao},
  doi     = {10.1016/j.physleta.2023.129103},
  journal = pla,
  pages   = {129103},
  title   = {{D}irect double ionization for molecules of {H}$_2$ and {N}$_2$ induced by electron-impact},
  volume  = {486},
  year    = {2023}
}

@article{gosd:liu23,
  author  = {Liu, Ya-Wei and Wang, Shu-Xing and Xu, Long-Quan and Peng, Yi-Geng and Zhang, Song-Bin and Wu, Y. and Wang, Jian-Guo and Zhu, Lin-Fan},
  doi     = {10.1103/PhysRevA.107.022803},
  journal = pra,
  pages   = {022803},
  title   = {{I}nelastic squared form factors of the vibronic states of ${B}^1{\Sigma}_\mathrm{u}^+$, ${C}^1{\Pi}_\mathrm{u}$, and ${EF}^1{\Sigma}_\mathrm{g}^+$ for molecular {D}$_2$ studied by high-energy electron scattering},
  volume  = {107},
  year    = {2023}
}

@article{gosd:toff24,
  author  = {Toffoli, D. and Coriani, S. and Stener, M. and Decleva, P.},
  doi     = {10.1016/j.cpc.2023.109038},
  journal = cpc,
  pages   = {109038},
  title   = {{T}iresia: {A} code for molecular electronic continuum states and photoionization},
  volume  = {297},
  year    = {2024}
}

@article{gosd:aker25,
  author  = {Aker, M. and others},
  collaboration = {The KATRIN Collaboration},
  doi     = {10.1126/science.adq9592},
  journal = science,
  pages   = {180--185},
  title   = {{D}irect neutrino-mass measurement based on 259 days of {KATRIN} data},
  volume  = {388},
  year    = {2025}
}

@article{gosd:naka25,
  author  = {Nakajima, I. and Yamazaki, M. and Popov, Y.~V. and Houamer, S. and Takahashi, M.},
  doi     = {10.1103/PhysRevA.111.052816},
  journal = pra,
  pages   = {052816},
  title   = {{T}esting reachability of the plane-wave impulse approximation in terms of the ratio of spectroscopic factors for $(e, 2e)$ ionization of {Ne} 2$s$ and 2$p$ electrons},
  volume  = {111},
  year    = {2025}
}

@article{gosd:roma26,
  author  = {Romano, D.~J. and Savitzky, B.~H. and Carrascosa, A.~M. and Narkiewicz-Jodko, A. and Geiser, J.~D. and Wang, P.~Y. and Huang, L. and Crane, S.~W. and Cheng, X. and Liang, M. and Mous, S. and Minitti, M.~P. and Simmermacher, M. and Kirrander, A. and Weber, P.~M.},
  doi     = {10.1039/D6FD00068A},
  journal = {Faraday Discuss.},
  pages   = {},
  volume  = {},
  title   = {{S}pectrally {R}esolved {X}-{R}ay {S}cattering},
  year    = {2026}
}

@article{gosd:volk26,
  author  = {Volkmann, H. and Saenz, A.},
  doi     = {10.1007/s00601-026-02061-8},
  journal = fbs,
  pages   = {38},
  title   = {{A} {D}istorted {S}ingle-{C}enter {A}pproach to {T}wo-{C}enter {C}oulomb {S}cattering},
  volume  = {67},
  year    = {2026}
}

@string{ADNDT="Atomic Data and Nuclear Data Tables"}

@string{ADP="Ann.\,der\,Phys."}

@string{AM="Ann.\,Math."}

@string{BCSJ="Bull.\,Chem.\,Soc.\,Jpn."}

@string{ChinPB="Chin. Phys. B"}

@string{CMP="Comm.\,Math.\,Phys."}

@string{CP="Chem.\,Phys."}

@string{CPC="Comp.\,Phys.\,Comm."}

@string{CPL="Chem.\,Phys.\,Lett."}

@string{EPJC="Eur.\,Phys.\,J.\,C"}

@string{EPJD="Eur.\,Phys.\,J.\,D"}

@string{FBS="Few-Body\,Syst."}

@string{IJQC="Int.\,J.\,Quant.\,Chem."}

@string{JCP="J.\,Chem.\,Phys."}

@string{JCTC="J.\,Chem.\,Theory Comput."}

@string{JPB="J.\,Phys.\,B"}

@string{JPG="J.\,Phys.\,G"}

@string{MP="Mol.\,Phys."}

@string{NI="Nucl.\,Instr.\,Meth."}

@string{PCCP="Phys.\,Chem.\,Chem.\,Phys."}

@string{PLA="Phys.\,Lett.\,A"}

@string{PPS="Proc.\,Phys.\,Soc."}

@string{PR="Phys.\,Rev."}

@string{PRA="Phys.\,Rev.\,A"}

@string{PRB="Phys.\,Rev.\,B"}

@string{PRL="Phys.\,Rev.\,Lett."}

@string{PRP="Phys.\,Rep."}

@string{QAM="Quart.\,Appl.\,Math."}

@string{RMP="Rev.\,Mod.\,Phys."}

@string{RPC="Rad.\,Phys.\,Chem."}

@string{RPP="Rep.\,Prog.\,Phys."}

@string{Science="Science"}

@string{ZNA="Z.\,Naturf.\,A"}

@string{ZP="Z.\,Phys."}

\end{document}